\newcounter{MYtempeqncnt}
\documentclass[10pt,journal,a4paper,final,oneside,twocolumn]{IEEEtran}

\usepackage[utf8]{inputenc}
\usepackage{kbordermatrix}  
\usepackage{blkarray}

\usepackage{times} 
\usepackage{graphicx}
\usepackage{cite}
\usepackage{citesort}
\usepackage{color}
\usepackage{psfrag}
\usepackage{subfigure}
\usepackage{amssymb}
\usepackage{epsfig}
\usepackage{pifont}
\usepackage{amsmath}  
\usepackage{array}
\usepackage{multicol}
\usepackage[T1]{fontenc}

\renewcommand{\baselinestretch}{0.99}
\title{Molecular Code-Division Multiple-Access: Signaling, Detection, and Performance}

\author{Weidong Gao, {\em Member,
    IEEE}, Lu Shi, {\em Member,
    IEEE}, and Lie-Liang Yang, {\em Fellow,
    IEEE} \thanks{W. Gao is with the College of Electronic Information Engineering, Wuxi University, Wuxi, 214105, China (Email: wdgao@cwxu.edu.cn); L. Shi is with the School of Intelligent Technology and Engineering, Chongqing University of Science and Technology, Chongqing, 401331, China (Email: 2020066@cqust.edu.cn); L.-L. Yang is with the School of Electronics
    and Computer Science, University of Southampton, SO17 1BJ,
    UK. (Email: lly@ecs.soton.ac.uk,~http://www-mobile.ecs.soton.ac.uk/lly).   
    ~~}}

\begin{document}

\maketitle

\begin{abstract} 
To accomplish relatively complex tasks, in Internet of Bio-Nano Things (IoBNT), information collected by different nano-machines (NMs) is usually sent via multiple-access channels to fusion centers (FCs) for further processing. Relying on two types of molecules, in this paper, a molecular code-division multiple-access (MoCDMA) scheme is designed for multiple NMs to simultaneously send information to an access-point (AP) in a diffusive molecular communications (DMC) environment. We assume that different NMs may have different distances from AP, which generates  `near-far' effect. Correspondingly, the uniform and channel-inverse based molecular emission schemes are proposed for NMs to emit information molecules. To facilitate the design of different signal detection schemes, the received signals by AP are represented in different forms. Specifically, by considering the limited computational power of nano-machines, three low-complexity detectors are designed in the principles of matched-filtering (MF), zero-forcing (ZF), and minimum mean-square error (MMSE). The noise characteristics in MoCDMA systems and the complexity of various detection schemes are analyzed. The error performance of the MoCDMA systems with various molecular emission and detection schemes is demonstrated and compared. Our studies and performance results demonstrate that MoCDMA constitutes a promising scheme for supporting multiple-access transmission in DMC, while the channel-inverse based transmission can ensure the fairness of communication qualities (FoCQ) among different NMs. Furthermore, different detection schemes may be implemented to attain a good trade-off between implementation complexity and communication reliability.
\end{abstract}

\section{Introduction}

With the global outbreak of COVID-19, human's well-being has attracted intensive research interest of the researchers from various fields. Nanotechnologies involving the manufacture of nano-material and nano-devices have revolutionized the ways for disease diagnosis and treatment~\cite{KHAN2019908,9056855}. Though nano-devices like bio-sensors have the advantages of, such as, tiny size, energy saving, and biocompatibility, making them suitable for medical and healthcare applications, an individual nano-machine (NM) is usually incapable of completing complex tasks, due to its low-energy storage and low computation capability~\cite{GALAL201845,9273037}. To accomplish more complicated biomedical tasks, including targeted drug delivery~\cite{freitas2006pharmacytes,timko2010remotely}, health monitoring~\cite{9121312,8735961}, tissue engineering~\cite{li2003cholesterol}, and tumor diagnosis~\cite{pan2011swallowable}, etc, a recent revolution trend is to envision the concept of Internet of Things (IoT) at the nano-scale, which is referred to the Internet of Bio-Nano Things (IoBNT)~\cite{7060516,6122529,7114200,9467302}. {It has been recognized that IoBNT has the potential to fill the interaction gaps between living cells (or NMs) and the Internet, where it is expected to play an essential role~\cite{10065592}.} 

There are various candidate techniques for designing the nano-networks or IoBNT, which include molecular communication (MC)~\cite{6949069,6208883}, electromagnetic communication at terahertz band~\cite{5675779,akyildiz2010electromagnetic,10115286}, acoustic communication (AC), and mechanical communication, etc. However, among them, MC belongs to a bio-inspired communication paradigm and has the most biocompatibility~\cite{9321459}. It enables molecules to convey information from nano-transmitters, such as NMs/sensors, to an access point (AP), like a fusion centre, which is separated from the nano-transmitters with certain distances~\cite{DMC-Akyildiz:2008:NNC:1389582.1389830}. 

Since in nature, free diffusion is a common propagation mechanism~\cite{8742793}, molecular communication via diffusion, or diffusive molecular communication (DMC) can be facilitated by the movements of information-embedded molecules solely driven by the diffusion process in liquid or gas media. Owing to this, DMC systems are energy-efficient, which benefits the durability of IoBNT. In DMC systems, information can be conveyed via modifying the concentration levels of molecules at transmitter. Correspondingly, the receiver decodes information according to the concentration levels obtained from its detection space~\cite{DMC-ParcerisaGine:2009:MCO:1631878.1631954}. However, due to the random diffusive movements of molecules, DMC usually experiences long transmission delay and strong inter-symbol interference (ISI)~\cite{8761545}, in addition to the noise generated from the random movements of information molecules as well as from the other noisy sources in the communications environment~\cite{Channel-5713270}.    

To achieve collaboration among individual NMs, a significant challenge in IoBNT is to enable multiple NMs to share a common DMC medium, namely multiple-access DMC, in order for them to send information to an AP for further processing or fusion~\cite{6208883}.
As conventional radio-based wireless communications, in principle, multiple-access DMC can be implemented in different domains. The most straightforward one is the molecular division multiple-access (MDMA)~\cite{DMC-ParcerisaGine:2009:MCO:1631878.1631954,GURSOY201845}, which is in parallel with the frequency division multiple-access (FDMA) in wireless communications. With MDMA, different NMs simultaneously communicate with an AP via orthogonal channels consisting of different types of information molecules. MDMA is simple {but the ratio between the number of molecular types and the number of NMs has to be an integer. To remove this limit, the molecular type spread molecular shift-keying (MTS-MoSK) scheme was proposed~\cite{10088444,ITUJournal}. Aided by the different spreading patterns, in MTS-MoSK, a same set of molecular types is used by all NMs to transmit information to an AP.} However, when there are many NMs, many types of molecules are required, which is highly challenging in terms of the synthesis and storage of molecules as well as the design of nano-transceivers. For instance, as there are many types of molecules, the receiver at AP is required to be equipped with many types of receptors to sense these different types of molecules, which may demand a high complexity. 

Multiple-access DMC can also be achieved in the time domain using molecular time-division multiple-access (MoTDMA)~\cite{Suzuki2012MultiobjectiveTO,6784527,9076283}. To implement MoTDMA, the time axis can be divided into frames, and a frame is further divided into a number of time slots. In this way, messages of different NMs can be delivered to AP using only one type of molecules via scheduling different NMs to transmit on different time slots. It is well-known that a crucial requirement for implementing TDMA is that all the transmitted signals should be well synchronized. However, this is often challenging to achieve in DMC, unless all the transmit NMs have the same distance from their AP. This is because in DMC, the channel impulse response (CIR) is non-linear, which is highly sensitive to the transmission distance, making both the amplitude and width of the received pulses from different nano-transmitters very different, if they have different distances from AP~\cite{OOK-6708553}. 

Multiple-access DMC may also be achieved via space division multiple-access, forming the MoSDMA, which distinguishes different transmit NMs based on their CIR shapes~\cite{6503794,6655190}. However, the implementation of MoSDMA relies very much on the instantaneous channels from different NMs, which are hard to estimate due to the multiple-access interference (MAI) among different NMs. {Evolving from the conventional MoSDMA, a modulation scheme referred to as the binary direction shift keying (BDSK) was proposed~\cite{9759486} to support multiple NMs. BDSK encodes one bit of information into two different pumping directions. Owing to this, in a BDSK-based system, adjacent transmitters can pump the same type of information molecules in two opposite azimuth angles that are directed to their corresponding receivers to mitigate MAI. However, this scheme demands the detailed design of transceivers, including the spherical transceivers of considerably larger than the transmission distance, the fixed pair of non-rotatable transceivers, and the reactive surfaces of transmitters and fully absorbing receivers. Because of these, the propagation links and states of the transceivers in BDSK-based DMC systems have to be stationary during communications.}

Multiple-access DMC can also be achieved in the principles of CDMA, forming the MoCDMA~\cite{7565087,7938035}. In MoCDMA, different NMs are assigned different signatures to spread messages, which can then be used by AP to de-spread and detect the information conveyed by the NMs. In~\cite{7994832}, a physical experiment on MoCDMA has been conducted. MoCDMA has the advantage that it requires only a small (typically $1$ or $2$) number of molecular types. Furthermore, depending on the DMC scenarios, MoCDMA may be operated in synchronous mode or asynchronous mode. It can even be operated in the scenarios where different NMs have different distances from their common receiver or/and transmit at different data rates. To be more specific, in \cite{7565087}, a MoCDMA scheme has been proposed on the basis of a single type of molecules using the on-off keying (OOK) modulation. Correspondingly, in this type of MoCDMA, the optical spreading sequences having their elemental values in the additive group of $\{0,1\}$ are used. Due to the on-off nature of signature sequences, the studies in \cite{7565087} show that the achievable error performance of this MoCDMA system is generally poor, even when the number of NMs supported is as low as $2$ for a spreading factor of $10$. By contrast, the authors in~\cite{7938035} have proposed a MoCDMA scheme based on two types of molecules with the binary molecule-shift keying (MoSK) modulation. The comparison between this MoSK-based MoCDMA with the OOK-based MoCDMA shows that the former one has the potential to significantly outperform the latter one. However, the studies in \cite{7938035} are very specific, which assume the Walsh codes for spreading and the adaptive threshold detection. Furthermore, the system model is also not general for use, as it assumes the same transmission distance from any nano-transmitters to AP. 

Against the background, in this paper, we consider a MoCDMA system where the transmission distances from NMs to AP may be different. For transmission at NMs, we compare two emission schemes, namely, the uniform emission and channel-inverse emission. Specifically, with the uniform emission, each NM emits the same number of molecules per symbol according to the required signal-to-noise ratio (SNR). By contrast, with the channel-inverse emission, the number of molecules emitted by a NM is adjusted according to its distance from AP. Explicitly, the channel-inverse emission scheme is capable of achieving the FoCQ among the `near-far' NMs. 
In our MoCDMA system, in order to implement the antipodal signaling, two types of molecules, which are assumed to be a pair of isomers with the same diffusion coefficient~\cite{6708565}, are employed. One type (Type-A) is for sending symbol $+1$, and the other types (Type-B) is for sending symbol $-1$. At the AP receiver, a passive observer is employed, which counts the numbers of molecules of both Type-A and Type-B, based on which the information received from different NMs is detected. We examine in detail the observations in our MoCDMA and show that the received signals can be represented in the forms that are fully equivalent to that in the conventional radio-based CDMA (RdCDMA) systems with binary phase shift keying (BPSK) modulation~\cite{yang2009multicarrier}.

Due to the slow diffusive propagation of molecules, MoCDMA systems may suffer from different types of interference, including, inter-chip interference (ICI), inter-symbol interference (ISI), and multiple-access interference (MAI), as well as the background noise that increases with the number of molecules emitted in the system. Therefore, in MoCDMA systems, the AP receiver should be carefully designed to efficiently suppress the above-mentioned interference. However, the transceiver design in nano-network is usually constrained by a tight computation budget, only allowing relatively low-complexity techniques. 
{Along with MC, there are some equalizers, drawn from the conventional radio-based communication receivers, having been proposed to mitigate ISI. Specifically, a time-domain linear equalizer based on minimum mean-square error (MMSE) principle and a nonlinear decision-feedback equalizer (DFE) have been proposed for the DMC systems with OOK~\cite{6708551}. In~\cite{9380963}, the frequency-domain equalizers based on zero-forcing (ZF) and MMSE criteria have respectively been developed also for the DMC systems with OOK. 
While these equalizers are in general efficient and low-complexity, they have been  designed only for ISI mitigation in the point-to-point DMC systems with OOK modulation. By contrast, in this paper, we apply the ZF and MMSE principles to design detectors for the MoCDMA systems experiencing simultaneously ICI, ISI, MAI and background noise. Specifically, three low-complexity linear detection schemes are derived and studied, which are the matched-filtering (MF), ZF, and MMSE detectors.} To achieve low-complexity detection, all the three detectors implement the symbol-by-symbol detection. We analyze the complexity of the three detectors, and investigate and compare their error performance. Our numerical studies demonstrate the achievable performance and analyze the performance-complexity trade-off of the MoCDMA systems with two molecular emission strategies and three detection schemes.

In summary, the novelties and contributions of this paper are as follows:
\begin{itemize}

\item A MoCDMA system with binary MoSK (BMoSK) modulation is proposed to take the advantages of the well-developed signal processing techniques in the conventional RdCDMA systems with BPSK modulation, such as, signal representation, interference mitigation, signal detection, etc.

\item Different transmission distances between NMs and AP are assumed to make the DMC system model general and practical. Correspondingly, a synchronization scheme is proposed to align the expected maximum concentration points of the CIRs corresponding to different NMs.

\item Considering that the performance of MoCDMA systems is highly sensitive to the transmission distances of different NMs, in analogy with the power-control used in RdCDMA, two molecular emission schemes, namely, the uniform emission and channel-inverse emission, are proposed and compared. The performance results show that the channel-inverse emission is capable of improving the reliability of the NMs further away from AP. This allows the detection of the NMs at different locations to achieve a similar BER performance and, hence, mitigate the near-far problem.

\item Three linear detection schemes in the principles of MF, ZF, and MMSE are introduced for the signal detection in MoCDMA systems. The complexity of these detectors is analyzed and the performance of the MoCDMA systems with these detectors is investigated and compared. Our studies and performance results show that all these detection schemes are low-complexity detection schemes, while the ZF- and MMSE-detectors are capable of achieving more reliable detection than the MF-detector.

\end{itemize}

The rest of the paper is organized as follows. First, the system model, including transmitter, channel model, emission schemes, received signals, and research assumptions are stated in Section~\ref{System_Model}. The different representations of received signals are provided in Section~\ref{Subsection-Detection-III.A}. Section~\ref{Section-Detection} presents the derivation of three signal detection schemes. In addition, the complexities of different detectors are provided in Section~\ref{Section-Detection}. The bit error rate (BER) performance of the MoCDMA systems with different detectors and molecular emission schemes is compared in Section~\ref{section-Result}. Finally, the concluding summary is made in Section~\ref{Conclusion}.

\section{System Model of MoCDMA}\label{System_Model}

\begin{figure}[tb]
  \begin{center} 
   \includegraphics[width=0.99\linewidth]{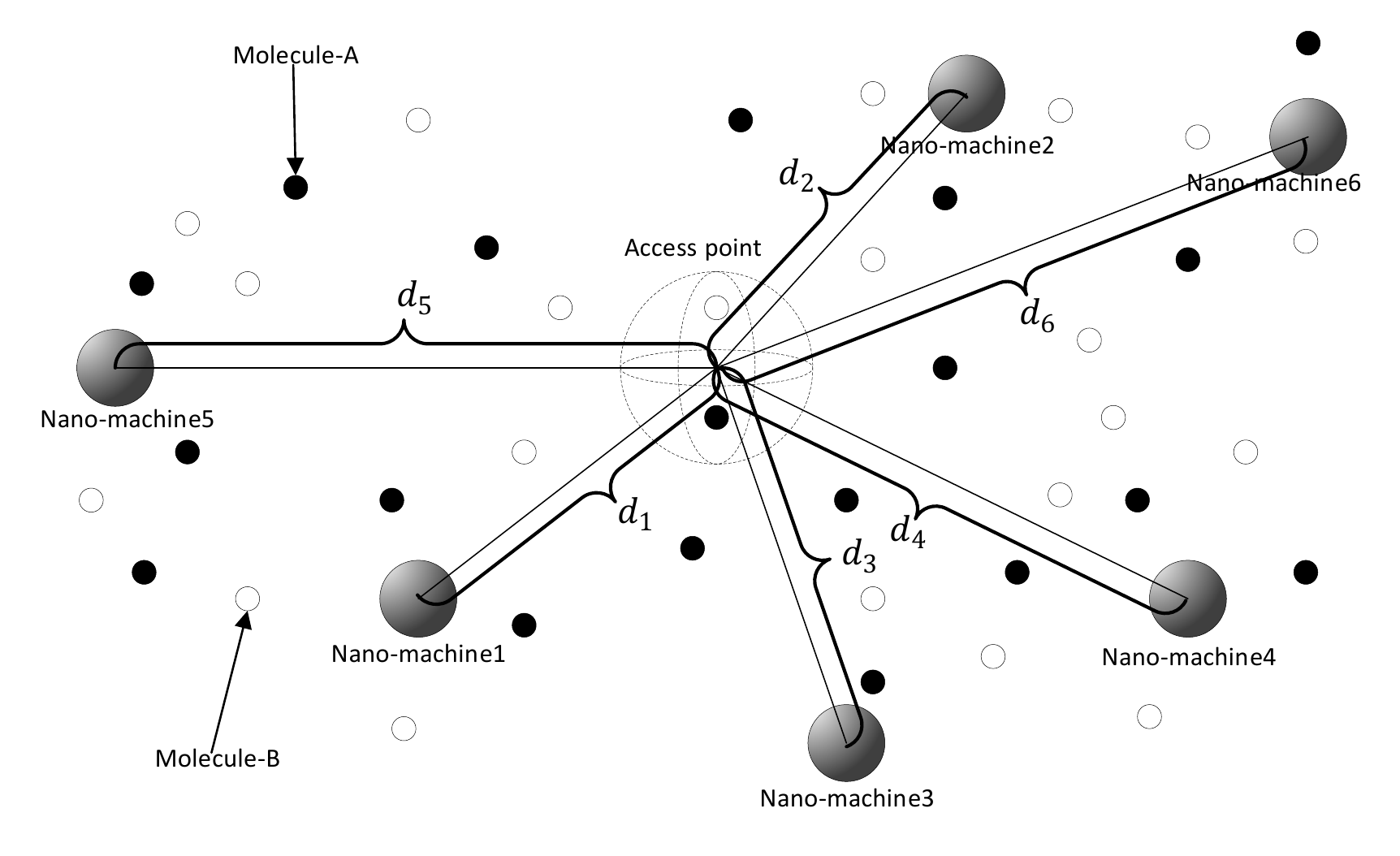}
   \end{center}
  \caption{System model for MoCDMA DMC systems, where nano-machines have different transmission distances to a common access point.}
  \label{figure-MUDMCS}
\end{figure}

We consider a multiple-access DMC system that consists of $K$ point nano-machines (NMs) (transmitters) and one spherical passive nano-type AP (receiver)
with a radius $\rho$, as illustrated in Fig.~\ref{figure-MUDMCS}. The locations of AP and all NMs are fixed. Unlike the assumption in~\cite{7938035} for the multiple-access MC systems, this paper takes different transmission distances from NMs into account. Without any loss of generality, the transmission distances from NMs to AP are sorted as $d_1\leq d_2\leq \ldots \leq d_K$. Because the concentration of information molecules at AP is highly sensitive to the transmission distance, two emission strategies will be introduced after the channel model description in this section. Here, we generally denote $Q^{(k)}$ as the number of molecules emitted by the $k$th NM for transmitting a symbol. Following the principle of CDMA, an information symbol is transmitted via emitting $N$ molecular impulses of each having $Q_c^{(k)}=Q^{(k)}/N$ molecules, where $N$ is referred to as the spreading factor or the number of chips within one symbol-duration of $T_s$ seconds. The interval between two adjacent impulses is referred to as the chip-duration of $T_c$ seconds. Hence, we have $T_s=NT_c$. When BMoSK modulation is employed, $T_s=T_b$, where $T_b$ is the bit-duration. Below we describe in detail the transmitter, channel model, emission strategies, and receiver in MoCDMA systems.       

\subsection{Transmitter}

We assume that in the considered MoCDMA system, information transmitted by the $k$th NM is expressed as $\pmb{B}_k=\{b_{k,0},b_{k,1},\dots,b_{k,j},\dots\}$, where $b_{k,j}
\in \{1,-1\}$. According to the principle of CDMA, the $u$th bit of NM $k$ is spread by a $N$ length spreading sequence~\cite{Ziemer_Peterson} denoted by $\pmb{s}_k=[s_{k,0},s_{k,1},\dots,s_{k,N-1}]^T$, where $s_{k,n}\in\{1,-1\}$. In our MoCDMA system, as NMs have different distances from AP, we define a common and fixed $T_s$ to allow synchronous transmission from  NMs to AP, which will become apparent in the forthcoming discourses. Moreover, for AP to obtain good observations of all NMs' transmitted signals, a time offset (delay) denoted by $T_{o}^{(k)}$ is introduced by the $k$th NM to emit its molecular pulses. Note that this time offset of an NM is dependent on its pulse peak presenting at AP, which will be detailed after describing the channel model. Then, when assuming that BMoSK
modulation is employed, $Q_c^{(k)}$ molecules per pulse
are emitted for the activated type (Type-A or Type-B) of molecules. Note that, Type-A and Type-B molecules are assumed to be a pair of isomers, whose diffusion coefficients are generally the same. Therefore, the transmitted signal of NM $k$ can be
expressed as
\begin{align}\label{eq:Transmitted_signal}
&s_{k}(t)=\nonumber\\
&\sum_{u=0,1,\ldots}\sum_{n=0}^{N-1}\left[\frac{1+b_{k,u}s_{k,n}}{2} Q_c^{(k)}
    \delta_A(t-uT_s-nT_c-T_{o}^{(k)})\right.\nonumber\\
    &\left.~~~~~~~~~+\frac{1-b_{k,u}s_{k,n}}{2} Q_c^{(k)}
    \delta_B(t-uT_s-nT_c-T_{o}^{(k)})\right], \nonumber\\ 
    &~~~~~~~~~k=1,2,\dots,K
\end{align}
where $\delta_A(t)$ and $\delta_B(t)$ denote the pulses of Type-A
and Type-B molecules, respectively, which are defined as
$\delta_A(0)=1$ and $\delta_B(0)=1$, and $\delta_A(t)=0$
and $\delta_B(t)=0$ for all $t\neq0$.

\subsection{Channel Model}

Assume that the diffusion channel environment is a fluid medium without flow, which is in its stable state during a session of communication. Hence, the diffusion coefficient $D$ of isomers is constant. Assumed that AP is able to measure the concentration of molecules inside the spherical detection space. Furthermore, assume that precise synchronization can be achieved between NMs and AP. Based on these assumptions, and when a pulse of $Q_c^{(k)}$ molecules is released by a NM at $t=T_o^{(k)}$, the concentration measured inside AP at
$t>T_o^{(k)}$ follows Fick's law, which can be expressed as~\cite{OOK-6708553}
\begin{align}\label{eq:Fick_Law}
c_k(t)=\frac{Q_c^{(k)}}{[4\pi D(t-T_o^{(k)})]^{\frac{3}{2}}}\exp\left[-{\frac{d_k^{2}}{4D(t-T_o^{(k)})}}\right], t>T_o^{(k)}
\end{align}
As shown in (\ref{eq:Fick_Law}), the molecular concentration $c_k(t)$ is a
time (T)-domain pulse function, the shape of which can be defined as the channel impulse response (CIR)~\cite{9321467} 
\begin{align}\label{eq:Channel_information}
h_k(t)=\frac{1}{[4\pi D(t-T_o^{(k)})]^\frac{3}{2}}\exp\left[-{\frac{d_k^{2}}{4D(t-T_o^{(k)})}}\right], t>T_o^{(k)}
\end{align}

It can be shown that in (\ref{eq:Channel_information}), when $d_k$ is given, the maximum value of CIR or pulse peak, can be attained at $t_d^{(k)}=d^2_k/6D$, which has a value of $h_{k,max}=\left(\frac{3}{2\pi
  e}\right)^{\frac{3}{2}}/{d_k^3}$. In principle, this extreme point is the best sampling point, which yields the highest power for signal detection. However, the pulse peak value $h_{k,\max}$ decreases with $d_k^3$, making DMC systems highly sensitive to transmission distance. Therefore, molecular emission strategies need to be designed for the MoCDMA systems to handle this `near-far' problem.

\subsection{Emission Control}\label{subsection-EmissionControl}

Let us denote the distance between the furthest NM and AP as $d_K$. This distance parameter is assumed to be known to AP and all NMs\footnote{In practice, AP is able to estimate its distance from a NM based on its measured CIR from the NM. Similarly, AP can send pilot molecular pulses for all NMs to measure their distances from the AP.}. In this case, the peak point of the furthest NM's CIR occurs at $t_d^{(K)}=d^2_K/6D$. Correspondingly, the maximum magnitude of the CIR is $h_{K,max}=\left(\frac{3}{2\pi
  e}\right)^{\frac{3}{2}}/{d_K^3}$. With the aid of the above information, to synchronize the peak points of the different NM's CIRs to the same time at AP, individual NMs emit their pulses according to their distances $d_k$ from AP. For instance, this can be achieved as follows. Before the start of NMs' transmissions, AP sends some pilot signals, such as, pulses of other type of molecules to NMs. Based on the measurements of pilots, each NM can identify its distance from AP and hence determine its emitting time, so that the peaks of the pulses sent by different NMs arrive at AP at the same time, as above-mentioned. In detail, assume that NMs are equipped with accurate clocks~\cite{6517979,6179346,6626319}. Then, at the beginning of the first chip, the emission time delay of NM $k$ can be obtained to be $T_o^{(k)}=t_d^{(K)}-t_d^{(k)}$, relative to the $K$th NM furthest from AP. In this way, the AP can sample for the maximum magnitudes of the CIRs of the different NMs.   

In terms of the numbers of molecules emitted by different NMs, two approaches with low complexity are proposed and will be compared. The first approach is the uniform emission, which allocates the same number of molecules to all NMs. Specifically, when denoting one-bit budget of molecules by $Q$, the number of molecules emitted by NM $k$ to transmit a chip pulse is
\begin{align}\label{eq:uniform_emission}
Q_c^{(k)}=Q/N.
\end{align}
However, this approach generates the 'near-far' problem, yielding that the CIR peak of a closer NM is higher than that of a further away NM.

Therefore, with the second approach, the channel-inverse based emission is implemented, which adjusts the number of molecules emitted by a NM according to its CIR. From \eqref{eq:Fick_Law} and \eqref{eq:Channel_information} we know that $c_{k,max}$ and $h_{k,max}$ are related by $c_{k,max}=Q_c^{(k)}h_{k,max}$. Then, with the channel-inverse based emission, NMs emit the numbers of molecules per chip to achieve
\begin{align}\label{eq:channel-inverse_concentration}
c_{1,max}=c_{2,max}=\ldots=c_{K,max}
\end{align}
where $c_{K,max}$ is obtained by assuming that the number of molecules emitted per chip by the furthest NM is $Q/N$. Then, using the above equalities, the number of molecules emitted per chip by NM $k$ can be found to be 
\begin{align}\label{eq:channel-inverse_emission}
Q_c^{(k)}=\frac{Q}{N}\cdot\frac{h_{K,max}}{h_{k,max}}=\frac{Q}{N}\cdot\left(\frac{d_{k}}{d_K}\right)^3\leq\frac{Q}{N}.
\end{align}

Compared to the uniform emission, the channel-inverse based emission can save a considerable number of molecules while different NMs have a similar detection reliability, in addition to mitigating the near-far problem.

\subsection{Receiver}\label{sec-receiver}

Assume that AP is able to identify the two types of molecules (Type-A and Type-B) without absorbing them. Then, when $K$ NMs emit molecular impulses in the form of
\eqref{eq:Transmitted_signal} to AP, the sampled observation difference between Type-A and Type-B molecules at $t$ corresponding to the $n$th chip,
$n=0,1,\dots\,N-1$, within the $u$th bit-duration can be expressed as
\begin{align}\label{eq:CDMA-Multi-Noise-concentration}
&z(t)=z_{A}(t)-z_{B}(t)= \nonumber\\
&\sum_{k=1}^{K}\sum_{j=0}^{uN+n} \frac{1+b_{k,\lfloor j/N\rfloor} s_{k,j\mathbin{\%}N}} {2}  \left[ c_{k}(t-jT_c)+n_{A,k,j}(t) \right]\nonumber\\
&-\sum_{k=1}^{K}\sum_{j=0}^{uN+n} \frac{1-b_{k,\lfloor j/N\rfloor} s_{k,j\mathbin{\%}N }} {2}  \left[ c_{k}(t-jT_c)+n_{B,k,j}(t)\right], \nonumber\\
&~~(uN+n)T_c+T_o^{(k)}\leq t< (uN+n+1) T_c+T_o^{(k)}
\end{align}
where $ \lfloor \cdot \rfloor$ is the floor operation, $j\mathbin{\%}N$ is the
reminder of $j/N$, $n_{A,k,j}(t)$ and $n_{B,k,j}(t)$ are the Brownian motion noise caused respectively by the Type-A and Type-B molecules emitted
by the NM $k$ within the $j$th chip duration~\cite{Channel-5713270}. To be more specific, according to~\cite{Channel-5713270,Channel-6906290}, when the average
number of molecules per pulse is sufficiently large, we can apply the Gaussian-approximation on $n_{A,k,j}(t)$ and $n_{B,k,j}(t)$, with the mean being zero and variance being
\begin{align}\label{eq:Noise-Variance}
\sigma_{A,k,j}^2(t)=&\frac{1+b_{k,\lfloor
      j/N\rfloor} s_{k,j\mathbin{\%}N }} {2V} c_{k}(t-jT_c),\nonumber\\ 
\sigma_{B,k,j}^2(t)=&\frac{1-b_{k,\lfloor
      j/N\rfloor} s_{k,j\mathbin{\%}N}} {2V} c_{k}(t-jT_c),
\end{align}
respectively, where $V=\frac{4}{3}\pi \rho^{3}$ is the volume of
the spherical detection space with a radius of $\rho$. For convenience of presentation,
these distributions are denoted as
$n_{A,k,j}(t)\sim\mathcal{N}\left(0,\sigma^2_{A,k,j}(t)\right)$ and
$n_{B,k,j}(t)\sim\mathcal{N}\left(0,\sigma^2_{B,k,j}(t)\right)$,
respectively.

As illustrated in \eqref{eq:Fick_Law}, after NM $k$ emits a pulse of molecules at $t=T_o^{(k)}$, the expected peak concentration of $c(t)$ at AP occurs at $t=t_d^{(K)}$, which is the same for all the $K$ NMs due to the emission control described in Section~\ref{subsection-EmissionControl}. Therefore, to detect the $u$th bits of $K$ NMs, AP samples for the concentrations at $t=(uN+n)T_c+t_d^{(K)}$, where $n=0,1,\dots,N-1$. Correspondingly, the observation difference between two types of molecules can be expressed as {\eqref{eq:CDMA-Multi-Sample} shown on the top of next page.}
\begin{figure*}[htb!]
\normalsize
\begin{align}\label{eq:CDMA-Multi-Sample}
Z_{u,n}&=z_{A}(t=(uN+n)T_c+t_d^{(K)})-z_{B}(t=(uN+n)T_c+t_d^{(K)})\nonumber\\
&=\sum_{k=1}^{K}\sum_{j=0}^{uN+n} \frac{1+b_{k,\lfloor j/N\rfloor} s_{k,j\mathbin{\%}N }} {2} \left[ c_{k} \Big((uN+n-j)T_c+t_d^{(K)}\Big)+n_{A,k,j} \Big((uN+n)T_c+t_d^{(K)} \Big) \right]\nonumber\\
&~~~~~~~~~-\sum_{k=1}^{K}\sum_{j=0}^{uN+n} \frac{1-b_{k,\lfloor j/N\rfloor} s_{k,j\mathbin{\%}N }} {2} \left[ c_{k} \Big((uN+n-j)T_c+t_d^{(K)} \Big)+n_{B,k,j} \Big((uN+n)T_c+t_d^{(K)} \Big) \right]\nonumber\\
&~~~~~~~~~~~u=0,1,\ldots;~n=0,1,\ldots,N-1
\end{align}
\hrulefill
\begin{align}\label{eq:CDMA-Multi-Sample-L}
Z_{u,n}&=\sum_{k=1}^{K}\sum_{j=\max\{0,uN+n-L\}}^{uN+n} \frac{1+b_{k,\lfloor j/N\rfloor} s_{k,j\mathbin{\%}N }} {2} \left[ c_{k} (uN+n-j)+n_{A,k,j}(uN+n) \right]\nonumber\\
&~~~~~~~~~-\sum_{k=1}^{K}\sum_{j=\max\{0,uN+n-L\}}^{uN+n} \frac{1-b_{k,\lfloor j/N\rfloor} s_{k,j\mathbin{\%}N }} {2} \left[ c_{k} (uN+n-j)+n_{B,k,j}(uN+n) \right]\nonumber\\
&~~~~~~~~~~~u=0,1,\ldots;~n=0,1,\ldots,N-1
\end{align}
\hrulefill
\setcounter{MYtempeqncnt}{\value{equation}}
\setcounter{equation}{11}
\begin{align}\label{eq:CDMA-Multi-Sample-Leq7}
Z_{u,n}=&\sum_{k=1}^{K}\sum_{i=0}^{\min\{L,uN+n\}} b_{k,\lfloor (uN+n-i)/N\rfloor} s_{k,(n-i)\mathbin{\%}N } \left[ c_{k}(i)+n_{k,uN+n-i}(uN+n) \right]\nonumber\\
=&\sum_{k=1}^{K}\sum_{i=0}^{\min\{L,uN+n\}} c_{k}(i) s_{k,(n-i)\mathbin{\%}N }b_{k,\lfloor (uN+n-i)/N\rfloor} +N_{u,n}\nonumber\\
&~~u=0,1,\ldots;~n=0,1,\ldots,N-1
\end{align}
\setcounter{equation}{\value{MYtempeqncnt}}
\hrulefill
\setcounter{MYtempeqncnt}{\value{equation}}
\setcounter{equation}{15}
\setcounter{MaxMatrixCols}{20}
\begin{align}\label{eq:widetilde_C}
\tilde{\pmb{C}}=
\begin{bmatrix}
c_1(0)  &c_1(1)  &\cdots  &c_1(L)  &0       &0       &\cdots  &0       &\cdots  &0       &0       &\cdots  &0\\
0       &0       &\cdots  &0       &c_2(0)  &c_2(1)  &\cdots  &c_2(L)  &\cdots  &0       &0       &\cdots  &0\\ 
\vdots  &\vdots  &\ddots  &\vdots  &\vdots  &\vdots  &\ddots  &\vdots  &\ddots   &\vdots  &\vdots  &\ddots  &\vdots\\  
0       &0       &\cdots  &0       &0       &0       &\cdots  &0       &\cdots  &c_K(0)  &c_K(1)  &\cdots  &c_K(L)
\end{bmatrix}^T
\end{align}
\setcounter{equation}{\value{MYtempeqncnt}}
\hrulefill
\end{figure*}
In \eqref{eq:CDMA-Multi-Sample}, the terms with index $j\neq 0$, which represent the pulses sent before the $(uN+n)$th impulse, impose inter-chip interference (ICI) on the current chip. However, from the characteristics of $c(t)$ we can know that ICI reduces significantly with time. Hence, in analysis and performance evaluation, we can set a sufficiently long ICI of $L$ chips, while ignoring the ICI imposed by the other pulses beyond $L$ chips. In this case, $Z_{u,n}$ in
\eqref{eq:CDMA-Multi-Sample} can be denoted in the form of {\eqref{eq:CDMA-Multi-Sample-L} shown on the top of next page,}
where $c_{k} (uN+n-j)$ corresponds to $c_{k} \left((uN+n-j)T_c+t_d^{(K)} \right)$
and $n_{X,k,j}(uN+n)$ corresponds to $n_{X,k,j} \left((uN+n)T_c+t_d^{(K)}\right)$ in \eqref{eq:CDMA-Multi-Sample}, $X$ represents $A$ or $B$.

In \eqref{eq:CDMA-Multi-Sample-L}, when the $j$th pulse of $k$th NM  is considered, we have either $\frac{1}{2}\left(1+b_{k,\lfloor j/N\rfloor}\right.$  \allowbreak $\left.s_{k,j\mathbin{\%}N }\right)=1$ and $\frac{1}{2}\left(1-b_{k,\lfloor j/N\rfloor} s_{k,j\mathbin{\%}N }\right)=0$ if $b_{k,\lfloor
    j/N\rfloor} s_{k,j\mathbin{\%}N }=1$, or $\frac{1}{2}\left(1+b_{k,\lfloor j/N\rfloor} s_{k,j\mathbin{\%}N }\right)=0$ and $\frac{1}{2}\left(1-b_{k,\lfloor j/N\rfloor} s_{k,j\mathbin{\%}N }\right)=1$ if $b_{k,\lfloor
    j/N\rfloor} s_{k,j\mathbin{\%}N }=-1$. Furthermore, $n_{A,k,j}(uN+n)$ for $b_{k,\lfloor j/N\rfloor}s_{k,j\%N}=1$ and $n_{B,k,j}(uN+n)$ for $b_{k,\lfloor j/N\rfloor}s_{k,j\%N}=-1$ have the same statistical properties. Hence, \eqref{eq:CDMA-Multi-Sample-L} can be expressed in an equivalent form of
\begin{align}\label{eq:CDMA-Multi-Sample-Leq}
Z_{u,n}&=\sum_{k=1}^{K}\sum_{j=\max\{0,uN+n-L\}}^{uN+n} b_{k,\lfloor j/N\rfloor} s_{k,j\mathbin{\%}N } \left[ c_{k}(uN+n-j)\right.\nonumber\\
&\left.+n_{k,j}(uN+n) \right]\nonumber\\
&~~u=0,1,\ldots;~n=0,1,\ldots,N-1
\end{align}
where $n_{k,j}(uN+n)$ follows the Gaussian distribution of 
$\mathcal{N}\left(0,\sigma^2_{k,j}(u,n)\right)$ with
$\sigma_{k,j}^2(u,n)=V^{-1} c_{k}(uN+n-j)$.

Let $i=uN+n-j$. Then, \eqref{eq:CDMA-Multi-Sample-Leq} can be represented in
a more convenient form of {\eqref{eq:CDMA-Multi-Sample-Leq7} shown on the top of next page.} In \eqref{eq:CDMA-Multi-Sample-Leq7}, $N_{u,n}=\sum_{k=1}^{K}\sum_{i=0}^{\min\{L,uN+n\}} b_{k,\lfloor
    \left(uN+n-i\right)/N\rfloor} s_{k,\left(n-i\right)\mathbin{\%}N }n_{k,uN+n-i}\\(uN+n)$, which is Gaussian distributed as $\mathcal{N}\left(0,\sigma^2\right)$ with
$\sigma^2=V^{-1}\sum_{k=1}^{K}\sum_{i=0}^{\min\{L,uN+n\}}c_k(i)$. We should note that the noise $N_{u,n}$ in \eqref{eq:CDMA-Multi-Sample-Leq7} is irrelevant to the information bits transmitted, which makes design and analysis more convenient.   

In MoCDMA systems, as \eqref{eq:CDMA-Multi-Sample-Leq7} demonstrates, transmitted signals suffer from ISI, ICI, and MAI between NMs. Hence, the detector of AP should be designed carefully to achieve reliable detection. Furthermore, as shown by $Z_{u,n}$ in \eqref{eq:CDMA-Multi-Sample-Leq7}, although the noise in MoCDMA systems with BMoSK modulation is not directly related to the transmitted information, any emitted molecule results in the increase of noise power, in addition to its role in conveying information. This is because the Brownian motions of molecules contribute to the noise in DMC systems, which is usually referred to as the counting noise. Counting noise leads to the unintentional perturbation to the concentration (or number of molecules) in detection sphere predicted by the Fick's law~\cite{Channel-5713270}. Thus, in MoCDMA systems, the noise power increases with the number of supported NMs via ISI, ICI, and MAI. This phenomenon  is very different from that in the radio-based communication systems, where environment noise is usually the additive white Gaussian noise (AWGN) that is independent of the transmitted signals. 

Having obtained the equivalent representation for observations, as shown in \eqref{eq:CDMA-Multi-Sample-Leq7}, next we represent the received signals in different forms in~\ref{Subsection-Detection-III.A} for easily deriving and explaining the detection schemes in \ref{Section-Detection}.

\section{Representation of Received Signals}\label{Subsection-Detection-III.A}

Considering the challenges of transceiver implementation in MoCDMA systems, in this paper, we only consider the symbol-by-symbol detectors. In this case, AP carries out detection based on a $N$-length observation vector corresponding to one bit, such as bit $u>1$. Then, it can be shown that from~\eqref{eq:CDMA-Multi-Sample-Leq7}, we have an expression of
\setcounter{MYtempeqncnt}{\value{equation}}
\setcounter{equation}{12}
\begin{align}\label{eq:CDMA-Multi-Sample-Leq11}
\pmb{z}_u=&\sum_{k=1}^K\pmb{C}_{k2}\pmb{S}_{k2}\pmb{b}_{k2}+\pmb{n}_u\nonumber\\
=&\sum_{k=1}^K\left(\pmb{C}_{k0}\pmb{s}_kb_{k,u}+\pmb{C}_{k-1}\pmb{s}_kb_{k,u-1}\right)+\pmb{n}_u
\end{align}
where $\pmb{z}_u=\left[Z_{u,0},Z_{u,1},\ldots,Z_{u,N-1}\right]^T$,
$\pmb{n}_u=\left[N_{u,0},N_{u,1},\ldots,N_{u,N-1}\right]^T$,
$\pmb{C}_{k2}$ is a $(N\times 2N)$ matrix whose last row is constituted
by zeros and $\textrm{Rev}(\pmb{c}^T_k)=[c_k(L),c_k(L-1),\ldots,c_k(0)]$, 
and $c_k(0)$ is located at the position of $(N-1,2N-1)$. In the other
$(N-1)$ rows, the parts of non-zero elements are just the shifts of 
these $(L+1)$ elements, and the left shift of the $i$th row 
is $(N-1-i)$ for $i=0,1,\ldots,N-2$. In \eqref{eq:CDMA-Multi-Sample-Leq11},
$\pmb{S}_{k2}=\pmb{I}_2\otimes \pmb{s}_k$, where $\pmb{I}_2$ is a $(2\times 2)$
identity matrix and $\otimes$ represents the Kronecker product operation, and
$\pmb{b}_{k2}=[b_{k,u-1},b_{k,u}]^T$. Note that, to obtain~\eqref{eq:CDMA-Multi-Sample-Leq11}, we assumed that the $u$th bit experiences interference only from a previous bit $(u-1)$. Correspondingly, the second equation in
\eqref{eq:CDMA-Multi-Sample-Leq11} explicitly represents the two bits separately, where $\pmb{C}_{k-1}$ consists of the first $N$ columns of $\pmb{C}_{k2}$, while
$\pmb{C}_{k0}$ is structured by the other $N$ columns of $\pmb{C}_{k2}$.

It can be shown that the second equation in
\eqref{eq:CDMA-Multi-Sample-Leq11} has another form of
\begin{align}\label{eq:CDMA-Multi-Sample-Leq12}
\pmb{z}_u=&\sum_{k=1}^K\left(\tilde{\pmb{S}}_{k,0}\pmb{c}_kb_{k,u}+\tilde{\pmb{S}}_{k,-1}\pmb{c}_kb_{k,u-1}\right)+\pmb{n}_u
\end{align}
where $\pmb{c}_k=[c_k(0),c_k(1),\ldots,c_k(L)]^T$, $\tilde{\pmb{S}}_{k,0}$ is a
$N\times (L+1)$ lower diagonal matrix structured as: a) $s_{k,0}$ is
on diagonal; b) the first column is occupied by $\pmb{s}_k$ and otherwise zeros; and c) the
other columns are constituted by the downshifts of the first column after
deleting the elements that are shifted outside the matrix. $\tilde{\pmb{S}}_{k,-1}$
is also a $N\times (L+1)$ matrix, which has only $L$ non-zero rows starting
from the top (row $0$) of the matrix. These non-zero rows are structured as follows: a) from left to right, the first row has the elements of $\{0,s_{k,N-1},s_{k,N-2}, \ldots, s_{k,N-L}\}$; and b) the other rows are obtained from right shifting this row by adding zeros to the left and deleting non-zero elements from the right, if they are shifted beyond the $L$th column. For instance, the second row is $\{0,0,s_{k,N-1},s_{k,N-2},\ldots,s_{k,N-L+1}\}$ and the final non-zero row is $\{0,0,\ldots,0,s_{k,N-1}\}$; and c) except the above, all the other locations are occupied by zeros.

In addition, when writing in a more compact form,
\eqref{eq:CDMA-Multi-Sample-Leq12} can be represented as
\begin{align}\label{eq:CDMA-Multi-Sample-Leq16}
\pmb{z}_u=&\tilde{\pmb{S}}_{0}\tilde{\pmb{C}}{\pmb{b}}_{u}+\tilde{\pmb{S}}_{-1}\tilde{\pmb{C}}\pmb{b}_{u-1}+\pmb{n}_u
\end{align}
where
$\tilde{\pmb{S}}_{0}=\left[\tilde{\pmb{S}}_{1,0},\tilde{\pmb{S}}_{2,0},\ldots,\tilde{\pmb{S}}_{K,0}\right]$, 
$\tilde{\pmb{S}}_{-1}=\left[\tilde{\pmb{S}}_{1,-1},\tilde{\pmb{S}}_{2,-1},\ldots,\tilde{\pmb{S}}_{K,-1}\right]$, 
$\pmb{b}_{u}=\left[b_{1,u},b_{2,u},\ldots,b_{K,u}\right]^T$, 
$\pmb{b}_{u-1}=\left[b_{1,u-1},b_{2,u-1},\ldots,b_{K,u-1}\right]^T$, and
$\tilde{\pmb{C}}$ is a $(LK\times K)$ matrix having the structure of {\eqref{eq:widetilde_C} shown on the top of the last page.}

With the various representations, as shown in \eqref{eq:CDMA-Multi-Sample-Leq11} -
\eqref{eq:CDMA-Multi-Sample-Leq16},  prepared, next we derive the detection schemes in Section~\ref{Section-Detection}.

\section{Signal Detection}\label{Section-Detection}

Due to the general computation limits in DMC, it is impractical to implement the detection schemes requiring high-complexity non-linear computation. Hence, our study focuses on three low-complexity linear detectors that are applicable for operation in DMC systems. Specifically, we consider the matched-filtering (MF) detection, also referred to as correlation detection, zero-forcing (ZF) detection, and minimum mean-square error (MMSE) detection.

It is well-known that for linear detectors, the decision variable for
the $u$th bit of the $k$th NM can in general be expressed as
\setcounter{MYtempeqncnt}{\value{equation}}
\setcounter{equation}{16}
\begin{align}\label{eq:CDMA-Multi-Sample-Leq17}
\varepsilon_{k,u}=\pmb{w}^T_{k,u}\pmb{z}_u,~u=0,1,\ldots;~k=1,2,\ldots,K
\end{align}
where $\pmb{w}_{k,u}$ is a weight vector, to be derived in the following subsections when different detection principles are respectively considered.

Based on the decision variable $\varepsilon_{k,u}$, the $u$th
bit of the $k$th NM is decided as
\begin{align}\label{eq:CDMA-Multi-Sample-Leq18}
\hat{b}_{k,u}=&\left\{\begin{array}{ll} \displaystyle 1, & \textrm{if
  $\varepsilon_{k,u}>0$}\\ \displaystyle 0~(\textrm{or $-1$}), & \textrm{else}
\end{array}\right., \nonumber\\
&u=0,1,\ldots;~k=1,2,\ldots,K.
\end{align}
Let us now describe the principles of different detection schemes in detail.

\subsection{Matched-Filtering Detection}\label{Subsubsection-Detection-IV.1.1}

The MF detection scheme is derived from \eqref{eq:CDMA-Multi-Sample-Leq12}. When employing MF-detector, the AP is not aimed at mitigating any interference but attempts to maximize the received energy with the aid of the knowledge about CIR and spreading sequence. Assume that AP attempts to detect the $u$th bit $b_{k,u}$ transmitted by a reference NM $k$ with the knowledge of spreading code $\pmb{s}_k$ and channel information $\pmb{c}_k$. According to the principle of MF, the weight vector can be formed as  
\begin{align}\label{eq:CDMA-Multi-Sample-Leq19}
\pmb{w}_{k,u}=\tilde{\pmb{S}}_{k,0}\pmb{c}_k.
\end{align}
Upon substituting it into \eqref{eq:CDMA-Multi-Sample-Leq17}, the decision variable $\varepsilon_{k,u}$ can be expressed as 
\begin{align}\label{eq:CDMA-Multi-Sample-Leq20}
\varepsilon_{k,u}=&\pmb{w}^T_{k,u}\pmb{z}_u\nonumber\\
=& \|\tilde{\pmb{S}}_{k,0}\pmb{c}_k\|^2b_{k,u} +\sum_{l\neq k}^K\pmb{c}_k^T\tilde{\pmb{S}}^T_{k,0}\tilde{\pmb{S}}_{l,0}\pmb{c}_kb_{l,u}\nonumber\\
&+\sum_{l=1}^K\pmb{c}_k^T\tilde{\pmb{S}}^T_{k,0}\tilde{\pmb{S}}_{l,-1}\pmb{c}_kb_{l,u-1}+\pmb{c}_k^T\tilde{\pmb{S}}^T_{k,0}\pmb{n}_u.
\end{align}
On the right-hand side of~\eqref{eq:CDMA-Multi-Sample-Leq20}, the first term is the desired output, the second term represents the MAI imposed by the other $(K-1)$ interfering NMs, the third term is the ISI by the previous $(u-1)$th
bits transmitted by all the $K$ NMs, and the final term is noise. 

In the case that the channel information $\pmb{c}_k$ is not available to AP, $\pmb{c}_k$ in \eqref{eq:CDMA-Multi-Sample-Leq19} can be replaced by
a $(L+1)$-length all-one vector expressed as $\pmb{1}$, yielding 
\begin{align}\label{eq:CDMA-Multi-Sample-Leq21}
\pmb{w}_{k,u}=\tilde{\pmb{S}}_{k,0}\pmb{1}.
\end{align}
Correspondingly, we have 
\begin{align}\label{eq:CDMA-Multi-Sample-Leq22}
\varepsilon_{k,u}=&  \pmb{1}^{T} \tilde{\pmb{S}}_{k,0}^{T} \tilde{\pmb{S}}_{k,0} \pmb{c}_kb_{k,u} +\sum_{l\neq k}^K\pmb{1}^T\tilde{\pmb{S}}^T_{k,0}\tilde{\pmb{S}}_{l,0}\pmb{c}_kb_{l,u}\nonumber\\
&+ \sum_{l=1}^K\pmb{1}^T\tilde{\pmb{S}}^T_{k,0}\tilde{\pmb{S}}_{l,-1}\pmb{c}_kb_{l,u-1}+\pmb{1}^T\tilde{\pmb{S}}^T_{k,0}\pmb{n}_u.
\end{align}

Referring to some concepts used in the radio-based communications~\cite{Proakis-5th}, the detector of \eqref{eq:CDMA-Multi-Sample-Leq20} carries out the maximal ratio combining (MRC), which is referred to as the MRC-detector, while \eqref{eq:CDMA-Multi-Sample-Leq22} executes the equal-gain combining (EGC), which is referred to as the EGC-detector for convenience. 

The MRC-detector attempts to maximize the signal-to-noise ratio
(SNR), while treating all interference as Gaussian noise, as shown
below. Upon substituting $\pmb{z}_u$ from
\eqref{eq:CDMA-Multi-Sample-Leq12} into
\eqref{eq:CDMA-Multi-Sample-Leq17}, the decision variable can be formed as
\begin{align}\label{eq:CDMA-Multi-Sample-Leq23}
\varepsilon_{k,u}=&\pmb{w}^T_{k,u}\left(\sum_{k=1}^K\left(\tilde{\pmb{S}}_{k,0}\pmb{c}_kb_{k,u}+\tilde{\pmb{S}}_{k,-1}\pmb{c}_kb_{k,u-1}\right)+\pmb{n}_u\right)\nonumber\\
=&\pmb{w}^T_{k,u}\tilde{\pmb{S}}_{k,0}\pmb{c}_kb_{k,u}+n_{k,u}
\end{align}
where $n_{k,u}=\pmb{w}^T_{k,u}\Big(\sum_{l\neq
  k}^K\tilde{\pmb{S}}_{l,0}\pmb{c}_kb_{l,u}+\sum_{l=1}^K\tilde{\pmb{S}}_{l,-1}\pmb{c}_kb_{l,u-1}
  +\pmb{n}_u\Big)$ is the weighted sum of interference and noise. When approximating this weighted sum as a random variable following Gaussian distribution of $\mathcal{N}(0,\pmb{w}^T_{k,u}\pmb{R}_I\pmb{w}_{k,u})$, where
$\pmb{R}_I$ is the autocorrelation matrix of interference-plus-noise,
i.e., of $\pmb{I}_n=\sum_{l\neq
  k}^K\tilde{\pmb{S}}_{l,0}\pmb{c}_kb_{l,u}+\sum_{l=1}^K\tilde{\pmb{S}}_{l,-1}\pmb{c}_kb_{l,u-1}+\pmb{n}_u$,
the SNR for detection of $b_{k,u}$ can be expressed as
\begin{align}\label{eq:CDMA-Multi-Sample-Leq24}
\gamma_{k,u}=&\frac{\|\pmb{w}^T_{k,u}\tilde{\pmb{S}}_{k,0}\pmb{c}_k\|^2}{\pmb{w}^T_{k,u}\pmb{R}_I\pmb{w}_{k,u}}\nonumber\\
=&\frac{\|(\pmb{R}_I^{1/2}\pmb{w}_{k,u})^T(\pmb{R}_I^{-1/2}\tilde{\pmb{S}}_{k,0}\pmb{c}_k)\|^2}{\pmb{w}^T_{k,u}\pmb{R}_I\pmb{w}_{k,u}}\nonumber\\
\leq & \frac{\|(\pmb{R}_I^{1/2}\pmb{w}_{k,u})^T\|^2\|(\pmb{R}_I^{-1/2}\tilde{\pmb{S}}_{k,0}\pmb{c}_k)\|^2}{\pmb{w}^T_{k,u}\pmb{R}_I\pmb{w}_{k,u}}
\end{align}
where `$\leq$' is satisfied due to the Cauchy-Schwarz
inequality~\cite{book:xiaodong-wang,book:Van-Trees}. It is well-known
that the equality in \eqref{eq:CDMA-Multi-Sample-Leq24} is satisfied,
if and only if
\begin{align}\label{eq:CDMA-Multi-Sample-Leq25}
\pmb{R}_I^{1/2}\pmb{w}_{k,u}=\alpha\pmb{R}_I^{-1/2}\tilde{\pmb{S}}_{k,0}\pmb{c}_k
\end{align}
where $\alpha$ is a constant. Hence, 
\begin{align}\label{eq:CDMA-Multi-Sample-Leq26}
\pmb{w}_{k,u}=&a\pmb{R}_I^{-1}\tilde{\pmb{S}}_{k,0}\pmb{c}_k,\nonumber\\
&~u=0,1,\ldots,M-1;~k=1,2,\ldots,K.
\end{align}

The detector with the weight vector of \eqref{eq:CDMA-Multi-Sample-Leq26} is capable of maximizing the signal to interference-plus-noise ratio (SINR), after mitigating the interference in $\pmb{I}_n$~\cite{6708553}. However, computing \eqref{eq:CDMA-Multi-Sample-Leq26} requires the knowledge about the spreading sequences employed by all NMs, and the CIRs $\pmb{c}$ of all NMs. If the receiver only has the knowledge about the reference NM to be detected, i.e., only knows $\pmb{s}_k$ (or $\tilde{\pmb{S}}_{k,0}$) and $\pmb{c}_k$, the detector is reduced to a MRC-detector. In this case, the detector has to approximate $\pmb{I}_n$ as a Gaussian noise vector having a covariance matrix of $\sigma_I^2\pmb{I}_N$. Consequently, the weight vector of \eqref{eq:CDMA-Multi-Sample-Leq26} is reduced to
\begin{align}\label{eq:CDMA-Multi-Sample-Leq27}
\pmb{w}_{k,u}=\frac{a}{\sigma^2_I}\tilde{\pmb{S}}_{k,0}\pmb{c}_k\equiv \tilde{\pmb{S}}_{k,0}\pmb{c}_k
\end{align}
which is \eqref{eq:CDMA-Multi-Sample-Leq19}. Therefore, the MRC-detector
maximizes SNR, after approximating interference to Gaussian noise.

\subsection{Zero-Forcing Detection}\label{Subsubsection-Detection-IV.1.2}

Based on \eqref{eq:CDMA-Multi-Sample-Leq16}, the ZF-detector can be readily derived. The conditions for implementing ZF detection include the knowledge of the spreading sequences employed by the $K$ NMs, and their CIRs to AP, so that the AP can construct $\tilde{\pmb{C}}$ and $\tilde{\pmb{S}}_0$ seen in \eqref{eq:CDMA-Multi-Sample-Leq16}. If these requirements are satisfied, according to~\eqref{eq:CDMA-Multi-Sample-Leq16}, the decision variables for the $K$ NMs can be obtained in ZF principle as
\begin{align}\label{eq:CDMA-Multi-Sample-Leq30}
\pmb{\varepsilon}_u=&\left(\tilde{\pmb{S}}_{0}\tilde{\pmb{C}}\right)^{\dagger}\pmb{z}_u\nonumber\\
=&\left(\tilde{\pmb{C}}^T\tilde{\pmb{S}}^T_{0}\tilde{\pmb{S}}_{0}\tilde{\pmb{C}}\right)^{-1}\tilde{\pmb{C}}^T\tilde{\pmb{S}}^T_{0}\pmb{z}_u,~u=0,1,\ldots,M-1
\end{align}
where $\pmb{A}^{\dagger}$ denotes the pseudo-inverse of
$\pmb{A}$~\cite{book:Van-Trees}.  Upon substituting $\pmb{z}_u$ from
\eqref{eq:CDMA-Multi-Sample-Leq16} into
\eqref{eq:CDMA-Multi-Sample-Leq30}, we have
\begin{align}\label{eq:CDMA-Multi-Sample-Leq29}
\pmb{\varepsilon}_u=& \pmb{b}_{u}+\underbrace{\left(\tilde{\pmb{C}}^T\tilde{\pmb{S}}^T_{0}\tilde{\pmb{S}}_{0}\tilde{\pmb{C}}\right)^{-1}\tilde{\pmb{C}}^T\tilde{\pmb{S}}^T_{0}\left(\tilde{\pmb{S}}_{-1}\tilde{\pmb{C}}\pmb{b}_{u-1}+\pmb{n}_u\right)}_{\textrm{ISI and noise}}
\end{align}
Eq.~\eqref{eq:CDMA-Multi-Sample-Leq29} shows that MAI is
fully removed. However, there is still ISI. 

Specifically, the weight vector for detecting the $u$th bit of the $k$
NM is given by the $k$th column of
$\left(\tilde{\pmb{S}}_{0}\tilde{\pmb{C}}\right)^{\dagger}$ in
\eqref{eq:CDMA-Multi-Sample-Leq30}, which can be expressed in detail
as
\begin{align}\label{eq:CDMA-Multi-Sample-Leq32}
\pmb{w}_{k,u}=\left(\tilde{\pmb{S}}_{0}\tilde{\pmb{C}}\left(\tilde{\pmb{C}}^T\tilde{\pmb{S}}^T_{0}\tilde{\pmb{S}}_{0}\tilde{\pmb{C}}\right)^{-1}\right)(:,k),~k=1,2,\ldots,K
\end{align}
where $\pmb{A}(:,k)$ represents the $k$th column of
$\pmb{A}$.

\subsection{Minimum Mean-Square Error Detection}\label{Subsubsection-Detection-IV.1.3}

We first derive the MMSE-detector based on~\eqref{eq:CDMA-Multi-Sample-Leq12} to detect the information transmitted by a NM.

The MMSE-detector is designed to minimize the mean-square error (MSE) between the estimated data and the actual data, which can be obtained from the optimization problem of
\begin{align}\label{eq:CDMA-Multi-Sample-Leq33}
\pmb{w}_{k,u}=&\arg\min_{\pmb{w}}\left\{E\left[\|b_{k,u}-\varepsilon_{k,u}\|^2\right]\right\}\nonumber\\
=&\arg\min_{\pmb{w}}\left\{E\left[\|b_{k,u}-\pmb{w}^T\pmb{z}_u\|^2\right]\right\}.
\end{align}
Let us define the cost function as
\begin{align}\label{eq:CDMA-Multi-Sample-Leq34}
J_{k,u}(\pmb{w})=&E\left[\|b_{k,u}-\pmb{w}^T\pmb{z}_u\|^2\right]\nonumber\\
=&1-2\pmb{r}_{ku}^T\pmb{w}+\pmb{w}^T\pmb{R}_{\pmb{z}_u}\pmb{w}
\end{align}
where $\pmb{R}_{\pmb{z}_u}$ is the auto-correlation matrix of $\pmb{z}_u$, which
can be derived from \eqref{eq:CDMA-Multi-Sample-Leq12} and expressed as
\begin{align}\label{eq:CDMA-Multi-Sample-Leq35}
\pmb{R}_{\pmb{z}_u}=\sum_{k=1}^K\left(\tilde{\pmb{S}}_{k,0}+\tilde{\pmb{S}}_{k,-1}\right)\pmb{c}_k\pmb{c}_k^T\left(\tilde{\pmb{S}}_{k,0}+\tilde{\pmb{S}}_{k,-1}\right)^T+\sigma^2\pmb{I}_N.
\end{align}
In \eqref{eq:CDMA-Multi-Sample-Leq34}, $\pmb{r}_{ku}$ is the
cross-correlation between $\pmb{z}_u$ and $b_{k,u}$, which is
\begin{align}\label{eq:CDMA-Multi-Sample-Leq36}
\pmb{r}_{k,u}=\tilde{\pmb{S}}_{k,0}\pmb{c}_k.
\end{align}
Upon taking the derivatives of $J_{k,u}(\pmb{w})$ with respect to $\pmb{w}$
and equating the result to zero, we obtain
\begin{align}\label{eq:CDMA-Multi-Sample-Leq37}
\frac{\partial J_{k,u}(\pmb{w})}{\partial (\pmb{w})}=-2\pmb{r}_{k,u}+2\pmb{R}_{\pmb{z}_u}\pmb{w}=0
\end{align}
from which we obtain the optimum weight vector acheving MMSE
detection as
\begin{align}\label{eq:CDMA-Multi-Sample-Leq38}
\pmb{w}_{k,u}=&\pmb{R}_{\pmb{z}_u}^{-1}\pmb{r}_{k,u}\nonumber\\
=&\Bigg(\sum_{k=1}^K\left(\tilde{\pmb{S}}_{k,0}+\tilde{\pmb{S}}_{k,-1}\right)\pmb{c}_k\pmb{c}_k^T\left(\tilde{\pmb{S}}_{k,0}+\tilde{\pmb{S}}_{k,-1}\right)^T\nonumber\\
&+\sigma^2\pmb{I}_N\Bigg)^{-1}\tilde{\pmb{S}}_{k,0}\pmb{c}_k,\nonumber\\
&~u=0,1,\ldots,M-1;~k=1,2,\ldots,K.
\end{align}

From \eqref{eq:CDMA-Multi-Sample-Leq38}, it can be inferred that to implement MMSE-detector, AP requires the knowledge about the spreading sequences employed by all NMs and their CIRs to the AP, so that AP can construct the matrices/vectors as shown in the formula. Furthermore, AP requires to know the noise power denoted by $\sigma^2$. In this case, the MMSE-detector is capable of achieving the best trade-off between interference (including both MAI and ISI) suppression and noise mitigation. It is worth noting that due to the BMoSK modulation employed, the noise power can be calculated, as shown in~\eqref{eq:CDMA-Multi-Sample-Leq7}.

In the case that AP does not have the knowledge of the spreading sequences of the $K$ NMs, but only that of the desired NM, AP has to approximate $\pmb{R}_u$ as a constant diagonal matrix, making \eqref{eq:CDMA-Multi-Sample-Leq38} be
reduced to the weight vector of the MRC-detector. In practice, even when there is no knowledge about the spreading sequences of the $K$ NMs, AP may estimate $\pmb{R}_{\pmb{z}_u}$ using the received observations using the formula $\pmb{\bar{R}}_{\pmb{z}_u}=\frac{1}{M}\sum^M_{u=1}\pmb{z}_u\pmb{z}^T_u$. Provided that $M$ is sufficiently large, usually  $\pmb{\bar{R}}_{\pmb{z}_u}\approx\pmb{R}_{\pmb{z}_u}$ can be satisfied. 

The above MMSE detector detects one NM after another. We can also derive a MMSE detector based on~\eqref{eq:CDMA-Multi-Sample-Leq16}, allowing AP to detect all NMs simultaneously. In this case, let $\pmb{W}_u$ be the $(N\times K)$ weight matrix. Then, the decision variable vector is given by 
\begin{align}
\pmb{\varepsilon}_u=\pmb{W}^T\pmb{z}_u
\end{align}
where the weight matrix in the MMSE principle is obtained from solving the optimization problem:
\begin{align}
\pmb{W}_{u}=&\arg\min_{\pmb{W}}\left\{E\left[\|\pmb{b}_{u}-\pmb{\varepsilon}_{u}\|^2\right]\right\}\nonumber\\
=&\arg\min_{\pmb{W}}\left\{E\left[\|\pmb{b}_{u}-\pmb{W}^T\pmb{z}_u\|^2\right]\right\}    
\end{align}
where $\pmb{z}_u$ is given by~\eqref{eq:CDMA-Multi-Sample-Leq16}.

Let us define the cost function as
\begin{align}
J_{u}(\pmb{W})=E\left[\|\pmb{b}_{u}-\pmb{W}^T\pmb{z}_u\|^2\right].
\end{align}
Then, when assuming that data bits are uniform random variables, the covariance matrix of $\pmb{b}_u-\pmb{W}^T\pmb{z}_u$ can be expressed as
\begin{align}
\pmb{J}_{u}(\pmb{W})=\pmb{I}_K-\pmb{R}^T_{\pmb{z}_u\pmb{b}_u}\pmb{W}-\pmb{W}^T\pmb{R}_{\pmb{z}_u\pmb{b}_u}+\pmb{W}^T\pmb{R}_{\pmb{z}_u}\pmb{W}   
\end{align}
where $\pmb{R}_{\pmb{z}_u}=E\left[\pmb{z}_u\pmb{z}_u^T\right]$ and $\pmb{R}_{\pmb{z}_u\pmb{b}_u}=E\left[\pmb{z}_u\pmb{b}_u^T\right]$, which are given by
\begin{align}\label{eq:CDMA-MMSE-R_z_u}
\pmb{R}_{\pmb{z}_u}=&\tilde{\pmb{S}}_{0}\tilde{\pmb{C}}\tilde{\pmb{C}}^T\tilde{\pmb{S}}^T_{0}+\tilde{\pmb{S}}_{-1}\tilde{\pmb{C}}\tilde{\pmb{C}}^T\tilde{\pmb{S}}_{-1}^T+\sigma_n^2\pmb{I}_N,\nonumber\\
\pmb{R}_{\pmb{z}_u\pmb{b}_u^T}=&\tilde{\pmb{S}}_0\tilde{\pmb{C}}.
\end{align}
Note that, it can be shown that $\pmb{R}_{\pmb{z}_u}$ in \eqref{eq:CDMA-MMSE-R_z_u} is the same as $\pmb{R}_{\pmb{z}_u}$ in \eqref{eq:CDMA-Multi-Sample-Leq35}.

The optimum weight matrix $\pmb{W}_u$ can be obtained by differentiating the trace of $\pmb{J}_u(\pmb{W})$ with respect to $\pmb{W}$ and equating the result to zero, i.e.,
\begin{align}
\frac{\partial \textrm{Tr} (\pmb{J}_u(\pmb{W}))}{\partial (\pmb{W})}=-2\pmb{R}_{\pmb{z}_u\pmb{b}_u^T}+2\pmb{R}_{\pmb{z}_u}\pmb{W}=0
\end{align}
the solution of which is the optimal solution, given as
\begin{align}\label{eq:CDMA-MMSE-W_u}
\pmb{W}_u=\pmb{R}_{\pmb{z}_u}^{-1}\pmb{R}_{\pmb{z}_u\pmb{b}_u^T}.
\end{align}
Upon substituting the results in \eqref{eq:CDMA-MMSE-R_z_u} into \eqref{eq:CDMA-MMSE-W_u}, we obtain
\begin{align}\label{eq:CDMA-MMSE-W_u-Leq2}
\pmb{W}_u=\left(\tilde{\pmb{S}}_{0}\tilde{\pmb{C}}\tilde{\pmb{C}}^T\tilde{\pmb{S}}^T_{0}+\pmb{R}_I\right)^{-1}\tilde{\pmb{S}}_{0}\tilde{\pmb{C}}
\end{align}
where by definition, $\pmb{R}_I=\tilde{\pmb{S}}_{-1}\tilde{\pmb{C}}\tilde{\pmb{C}}^T\tilde{\pmb{S}}_{-1}^T+\sigma_n^2\pmb{I}_N$.

Furthermore, after applying the matrix inverse lemma, we can obtain another representation for $\pmb{W}_u$, which is
\begin{align}\label{eq:CDMA-MMSE-W_u-Leq3}
\pmb{W}_u=&\pmb{R}_I^{-1}\tilde{\pmb{S}}_{0}\tilde{\pmb{C}}\left(\tilde{\pmb{C}}^T\tilde{\pmb{S}}_{0}^T\pmb{R}_I^{-1}\tilde{\pmb{S}_0}\tilde{\pmb{C}}+\pmb{I}_K\right)^{-1},\nonumber\\
&u=1,2,\ldots,M.
\end{align}
Additionally, when there is no ISI, making $\pmb{R}_I=\sigma^2_n\pmb{I}_N$, it can be easily shown that \eqref{eq:CDMA-MMSE-W_u-Leq3} is reduced to
\begin{align}\label{eq:CDMA-MMSE-W_u-Leq4}
\pmb{W}_u=&\tilde{\pmb{S}}_{0}\tilde{\pmb{C}}\left(\tilde{\pmb{C}}^T\tilde{\pmb{S}}_{0}^T\tilde{\pmb{S}_0}\tilde{\pmb{C}}+\sigma_n^2\pmb{I}_K\right)^{-1},\nonumber\\
&u=1,2,\ldots,M.
\end{align}

\subsection{Complexity of Detection Schemes}\label{Subsection-Detection-IV.4}

Considering that the weight vector $\pmb{w}_{k,u}$ in symbol-by-symbol detection is fixed in one session of detection, the computation at the AP can be separated into two components, namely the preparation of $\pmb{w}_{k,u}$ and the process of detection. The complexities of these two components are investigated respectively. To describe the complexity of the different cases, an arithmetic operation of the individual elements in a vector or matrix is assumed to have the complexity of $\mathcal{O}(1)$. Additionally, the complexity is expressed in terms of one bit and one NM. 

As shown in Sections \ref{Subsubsection-Detection-IV.1.1}, \ref{Subsubsection-Detection-IV.1.2} and \ref{Subsubsection-Detection-IV.1.3}, the decision variables  for all detection schemes to detect one bit have the form of \eqref{eq:CDMA-Multi-Sample-Leq17}. Thus, once the $N$-length weight vector is constructed, the detection complexity of all detectors, i.e., the MF- (or MRC-, EGC-), ZF-, and MMSE-detectors, is the same and is $\mathcal{O}(N)$ per bit per NM. By contrast, the complexity of computing the weight vectors by different detectors may be different. First, for the MF-detector, $\pmb{w}_{k,u}$ is given by \eqref{eq:CDMA-Multi-Sample-Leq19} for the MRC-detector, and by \eqref{eq:CDMA-Multi-Sample-Leq21} for the EGC-detector, both of which have the complexity of $\mathcal{O}(N(L+1))$. For the ZF-detector, corresponding to \eqref{eq:CDMA-Multi-Sample-Leq16}, the weight vector is given by \eqref{eq:CDMA-Multi-Sample-Leq32}. The complexity to calculate the inverse of a $(K\times K)$ dimensional matrix is $\mathcal{O}(K^3)$~\cite{Golub-Matrix-Computation}. Hence, it can be readily known that the complexity for computing the weight vector of the ZF-detector based on \eqref{eq:CDMA-Multi-Sample-Leq32} is $\mathcal{O}(N(L+1)+2NK+K^2)$. Finally, the complexity for preparing the weight vector based on \eqref{eq:CDMA-Multi-Sample-Leq38} for the MMSE-detector can be similarly analyzed, which is $\mathcal{O}(N^3/K+2N^2+3N(L+1)+N/K)$. 

The complexity of the different detection schemes is summarized in Table \ref{tabel:complexity}. Given $K\leq N$, it is shown that computing the weight vector for MMSE-detector demands the highest complexity, followed by the ZF-detector. MF-detector has the lowest complexity. However, we should note that, provided that the positions of NMs do not change, the weight vectors are fixed and are not required to be updated. In this case, all the three types of detection schemes have a similar complexity.

\begin{table}
    \caption{Complexity of computing weight vectors and that of detection.} 
    \label{tabel:complexity}
\centering
    \begin{tabular}{| l | l | l |}
    \hline
    Detector  & Computing weight vector & Detection \\ \hline
    MF & $\mathcal{O}(N(L+1))$ & $\mathcal{O}(N)$  \\
    ZF & $\mathcal{O}(N(L+1)+2NK+K^2)$ & $\mathcal{O}(N)$ \\
    MMSE & $\mathcal{O}(N^3/K+2N^2+3N(L+1)+N/K)$ & $\mathcal{O}(N)$ \\\hline
    
    \end{tabular}

\end{table}

\section{Performance Results and Discussion}\label{section-Result}

\begin{table}
    \caption{Parameters for performance study} 
    \label{tabel:parameter}
\centering
    \begin{tabular}{| l | l |}
    \hline
    Parameter  & Value \\ \hline
    Bit duration $T_b$ & $0.06s$ or $0.03s$ \\
    ISI length $L$ & $10$ \\
    Length of (MLS) spreading sequence $N$ & $31$ \\
    Radius of AP sensing space $\rho$ & $0.4~\mu m$ \\
    Diffusion coefficient $D$ & $4.5\times10^{-9}~m^2/s$ \\
    Distance between AP and NM $1$ $d_1$ & $2.2~\mu m$ \\
    Distance between AP and NM $2$ $d_2$ & $2.4~\mu m$ \\
    Distance between AP and NM $3$ $d_3$ & $2.6~\mu m$ \\
    Distance between AP and NM $4$ $d_4$ & $2.8~\mu m$\\
    Distance between AP and NM $5$ $d_5$ & $3.3~\mu m$ \\
    Distance between AP and NM $6$ $d_6$ & $3.5~\mu m$ \\
    \hline
        
    \end{tabular}

\end{table}

In this section, we demonstrate the performance of the MoCDMA systems with different settings and detection schemes. In the following figures, when there is no specific notification, the default parameters are given in Table \ref{tabel:parameter}. Note that in our study, spreading sequences are generated from the maximum length sequences (MLS). However, as MLSs are periodic sequences, one of them can be obtained from another one with a certain number of shifts \cite{Ziemer_Peterson}. Hence, they cannot be directly applied to the MoCDMA experiencing severe ISI and ICI. This is because, if the number of NMs supported exceeds some upper limit, the ICI in the current symbol duration and the ISI on the following durations may be fully despread by another sequence assigned to another NM, generating high MAI. Based on the above observation and to prevent interference, we set the length of spreading sequences to $N=31$ for supporting maximum $K=6$ NMs, when MLSs are employed. This setting allows the $31$-length MLSs to support upto $6$ NMs without full overlapping, when $L=10$. Besides MLSs, the performance of the MoCDMA systems with gold sequences and Walsh codes~\cite{Ziemer_Peterson} is also investigated and compared.

\begin{figure*}[th]
  \centering
    \subfigure[MRC-detector]{\epsfig{figure=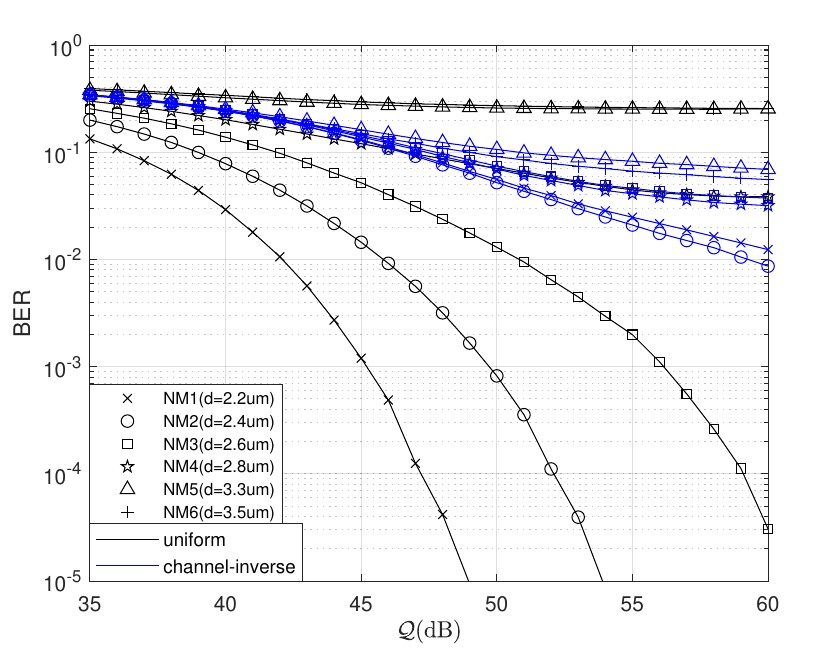,width=0.45\linewidth}}
    \subfigure[ZF-detector]{\epsfig{figure=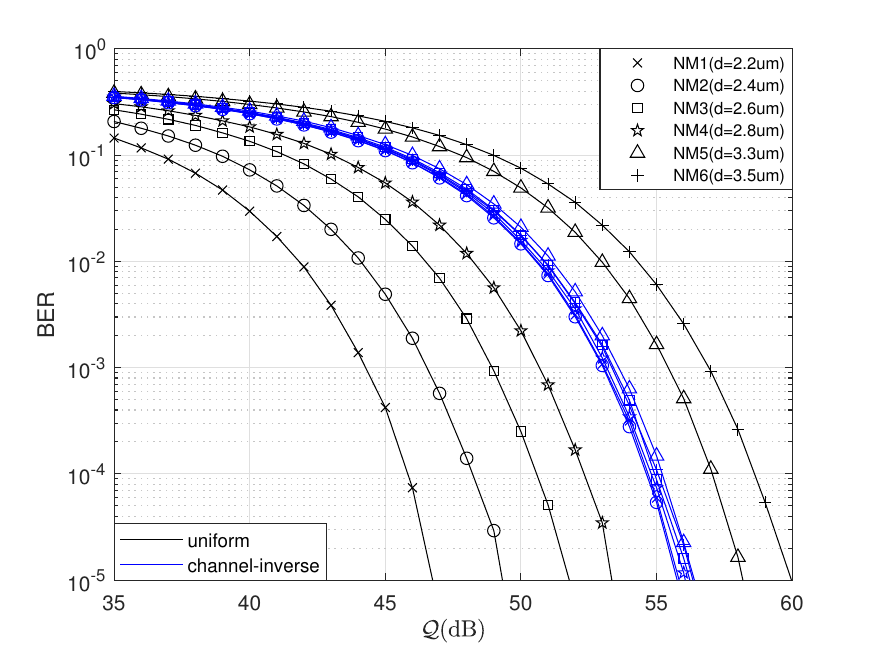,width=0.45\linewidth}} 
    \subfigure[MMSE-detector]{\epsfig{figure=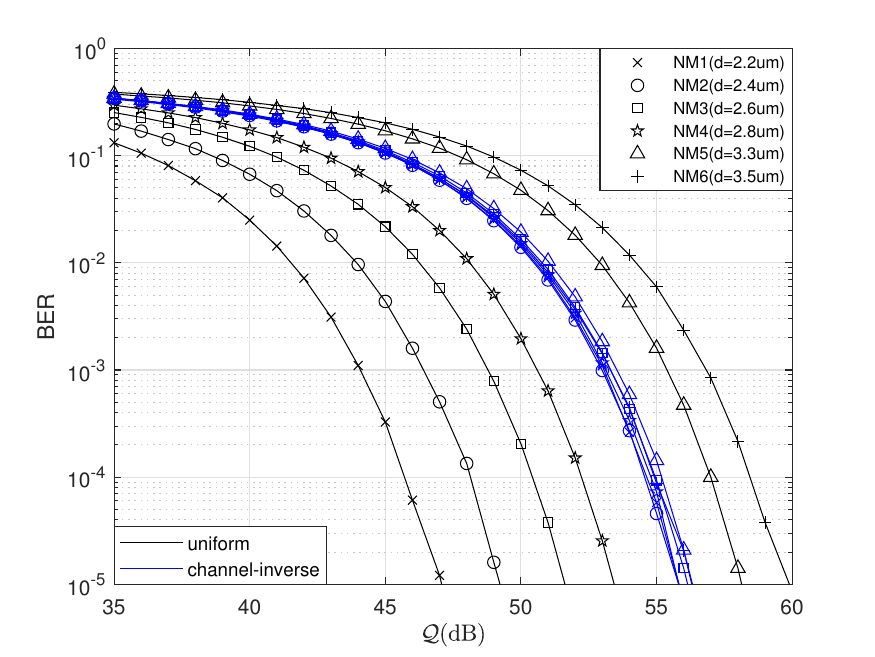,width=0.45\linewidth}}
    \subfigure[Real number of molecules emitted for one bit]{\epsfig{figure=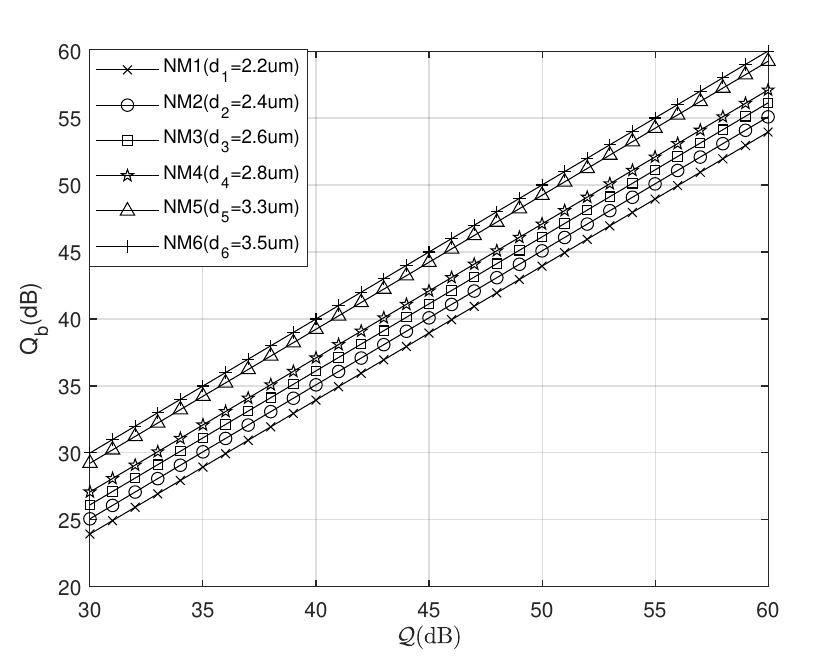,width=0.45\linewidth}}
  \caption{Comparison of BER performance of the MoCDMA systems with
    various detection schemes, when $T=6\times10^{-2}s$, $L=10$, $N=31$, and when two emission schemes are respectively considered.}
  \label{fig:BER_KQb}
\end{figure*}

Let us first compare the two emission control schemes addressed in
this paper in Fig.~\ref{fig:BER_KQb}, where all the detection schemes
described in Section~\ref{Section-Detection} are respectively
considered. The MoCDMA system considered in Fig.~\ref{fig:BER_KQb} supports $6$ NMs with the detailed settings listed in Table \ref{tabel:parameter}. It is worth noting that in Fig.~\ref{fig:BER_KQb}, $\mathcal{Q}$ represents the maximum number of molecules available to emit for one bit. As Section~\ref{subsection-EmissionControl} describes, in the case of uniform emission, all NMs emit the maximum number of molecules per bit. By contrast, in the case of the channel-inverse relied-on emission, NM $k$ emits the number of molecules per bit according to $Q_b^{(k)}\propto d_k^3$. Hence, in the channel-inverse emission scheme, the NMs relatively close to AP actually emit less molecules than $\mathcal{Q}$ as shown in Fig.~\ref{fig:BER_KQb}(d). Therefore, when a NM is closer to AP, it can save more molecules. Another advantage of the channel-inverse emission scheme is that, by reducing the number of emitted molecules, a NM closer to AP imposes lower interference on the NMs further away from AP, as well as reduces the power of counting (environment) noise.

As shown in Fig.~\ref{fig:BER_KQb}(a), when NMs adopt the uniform emission and AP employs MRC-detector, the distance between a NM and AP has a significant influence on the BER performance of MoCDMA systems. The communication reliabilities of the NMs with different distances from AP are in great differences. Specifically, for NMs $5$ and $6$, which are the furthest from AP, their information is completely undetectable regardless of the uniform or channel-inverse emission. By contrast, NM $1$, which is closest to AP, achieves a promising error performance that is only slightly lagging behind the corresponding one in Fig.~\ref{fig:BER_KQb}(c) with MMSE-detector. This is because the MRC-detector amplifies the impact of CIR, which causes a polarization among NMs. When the channel-inverse emission is employed, as shown in Fig.~\ref{fig:BER_KQb}(a), the BER performance of the far away NMs, NMs $5$ and $6$, is improved, while that of the NMs close AP like, NMs $1$ and $2$, is degraded. The reason behind is that the close AP NMs decrease their numbers of molecules emitted, which reduces MAI and counting noise and hence benefits the detection of the far away NMs. However, as the results in Fig.~\ref{fig:BER_KQb}(a) illustrate, the channel-inverse emission scheme in cooperation with the MRC-detection can hardly provide acceptable BER performance.

\begin{figure*}[h!]
  \centering
    \subfigure[ZF-detector]{\epsfig{figure=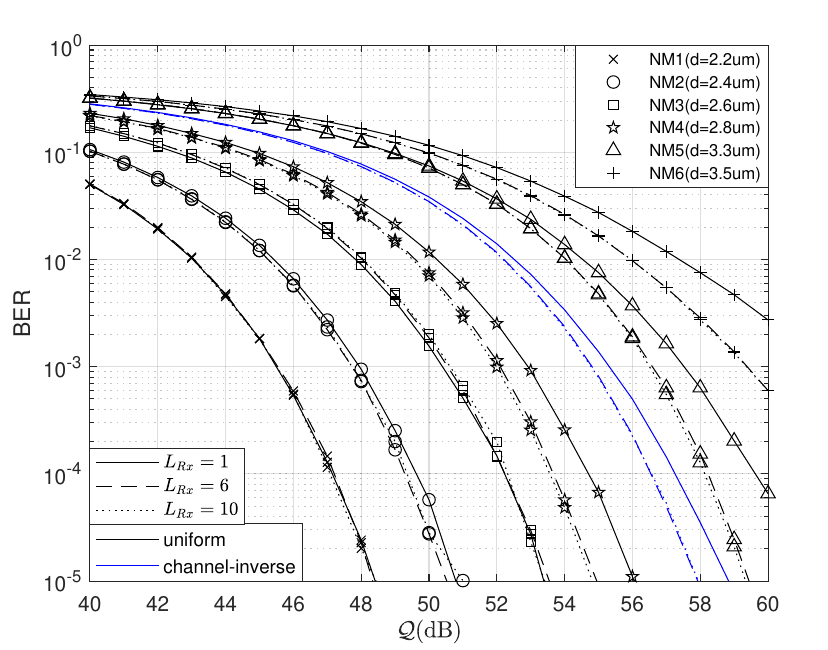,width=0.45\linewidth}}
    \subfigure[MMSE-detector]{\epsfig{figure=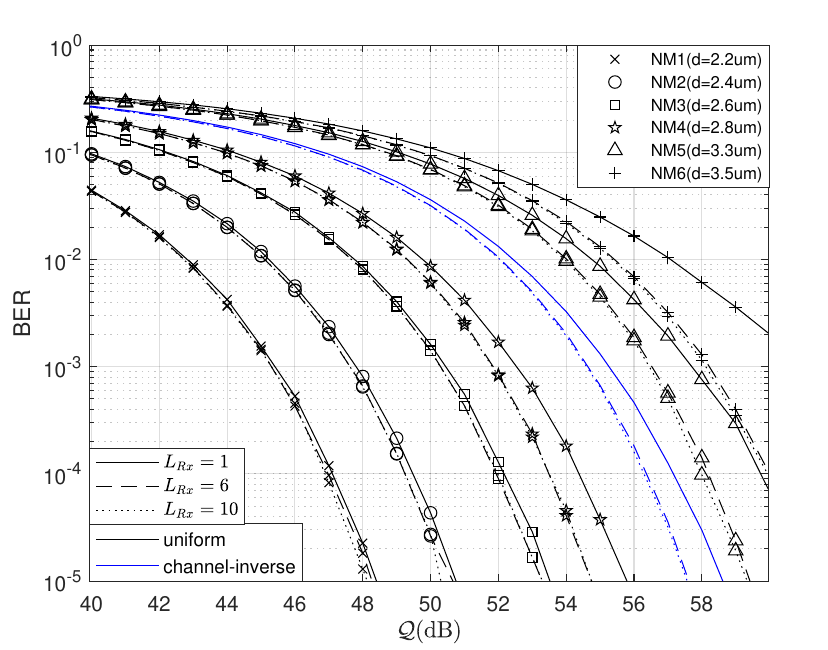,width=0.45\linewidth}} 
  \caption{Impact of the ISI-length exploited by AP on the BER performance of MoCDMA systems with ZF- and MMSE-detectors, when $T_b=0.03s$, $L=10$, $N=31$.}
  \label{fig:BER_RxLQb}
\end{figure*}

Compared to the MRC-detector, both the ZF-detector and MMSE-detector are highly efficient, when either the uniform or channel-inverse emission scheme is employed. As shown in Fig.~\ref{fig:BER_KQb}(b) and Fig.~\ref{fig:BER_KQb}(c), even under the uniform emission, the BER curves appear in the waterfall form, inferring that a low BER can be attainable, provided that SNR is sufficiently high. If the channel-inverse emission is adopted, the BER performance of all NMs with different transmission distances is similar, guaranteeing the FoCQ among NMs. This phenomenon can be explained as follows. First, the quality of MLSs assigned to different NMs is quite similar. Second, according to the principle of ZF-detector, it is capable of cancelling MAI while the channel-inverse emission provides the same received signal power for all NMs. Consequently, the post-processed SINR of all NMs can be nearly the same. Third, because the MMSE-detector can suppress interference and noise in the best trade-off, the most unfavorable factor can be mitigated. Therefore, AP can make the detection of all NMs at similar SINRs. In addition, a similar observation as that in Fig.~\ref{fig:BER_KQb}(a) can be obtained, i.e., although the channel-inverse emission results in some performance loss of the close AP NMs, it can improve the BER performance of the far away NMs. Finally, when comparing the ZF-detector with MMSE-detector, the MMSE-detector slightly outperforms the ZF-detector. 

With the second set of results shown in Fig.~\ref{fig:BER_RxLQb}, we demonstrate the impact of ISI length exploited by AP on the BER performance of the MoCDMA systems with respectively ZF- and MMSE-detectors, when $T_b=0.03s$ and the other parameters are shown in Table.~\ref{tabel:parameter}. We assume that the ISI length is $L=10$, and the ISI length exploited by AP is $L_{Rx}=1,~6$ or $10$. When the channel-inverses emission is used, due to the proximity of the BER performance of different NMs, the average BER of individual NMs is depcted as the blue-colored curves, instead of drawing the individual BER curves of all $6$ NMs. As the results in Fig.~\ref{fig:BER_RxLQb} show, for both the emission methods and both detectors, a better BER performance is generally achieved, as a longer ISI is exploited by AP for carrying out detection. Specifically, when the uniform emission is employed, the NMs located further away from AP achieve more explicit improvement in the BER performance, as the ISI length exploited increases. This is because as the distance increases, the received molecular concentration pulse becomes wider, resulting in that its peak value becomes similar to that of the residual concentration from previous transmissions, and hence, yielding severer interference and higher counting noise. In this case, exploiting more ISI by AP is helpful for the ZF- or MMSE-detector to mitigate interference and background noise.  However, as shown in Fig.~\ref{fig:BER_RxLQb}, provided that $L_{Rx}\geq 6$, the BER performance achieved by both the ZF- and MMSE-detectors, regardless of the uniform or channel-inverse emission, is nearly the same. This observation infers that although there is still ISI beyond $L_{Rx}=6$, the ISI is generally ignorable, as the result that the number of residual molecular is reduced to a very low level after about 6 chip durations.

\protect
\begin{figure*}[th]
  \centering
    \subfigure[ZF-detector (BTC)]{\epsfig{figure=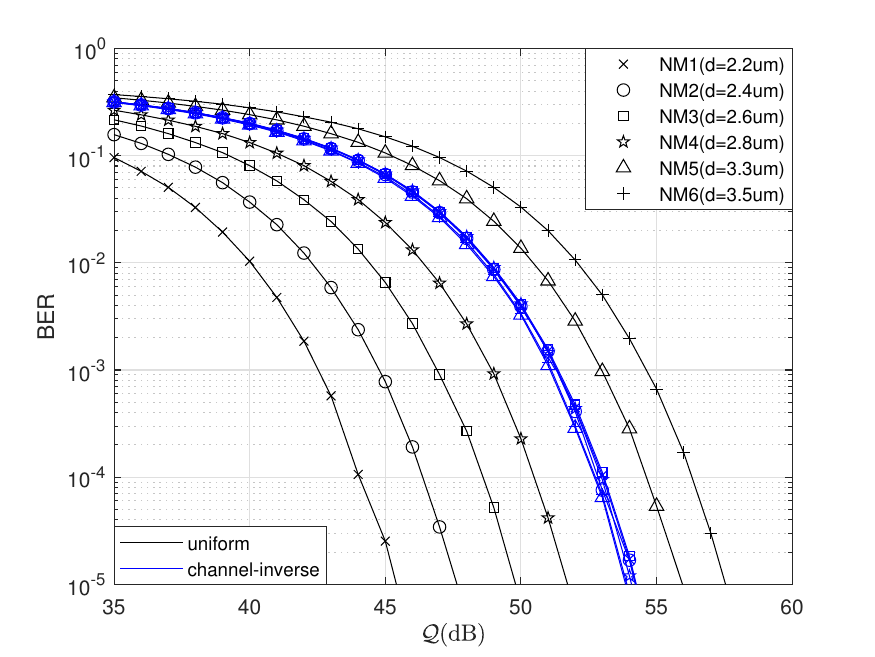,width=0.45\linewidth}}
    \subfigure[MMSE-detector (BTC)]{\epsfig{figure=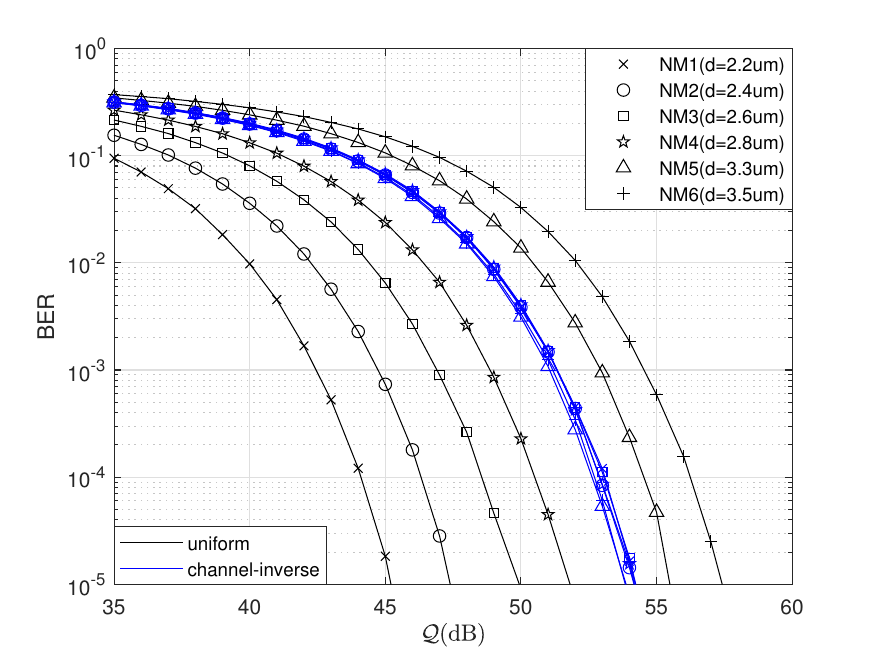,width=0.45\linewidth}}
    \subfigure[ZF-detector (BTF)]
    {\epsfig{figure=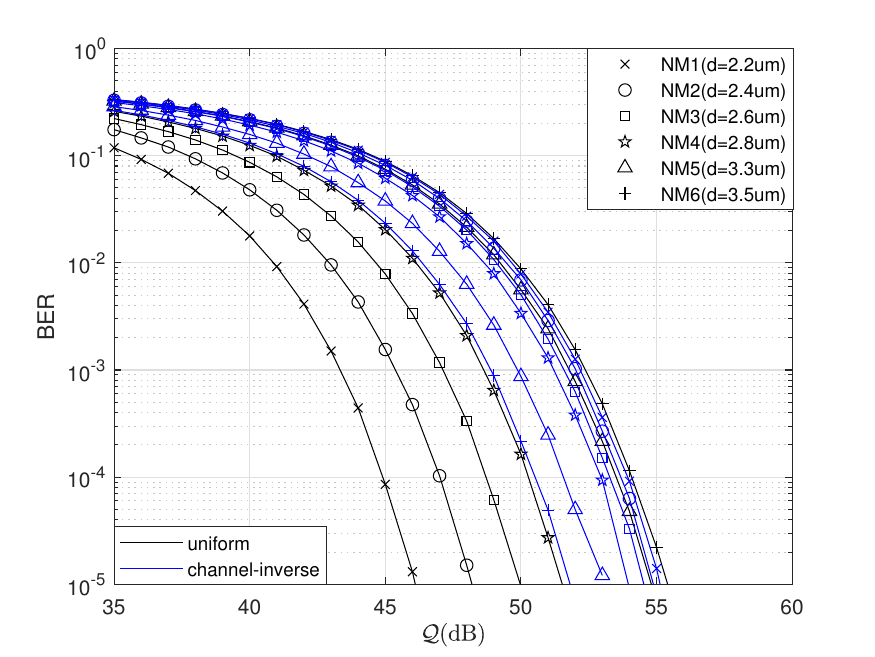,width=0.45\linewidth}}
     \subfigure[MMSE-detector (BTF)]
    {\epsfig{figure=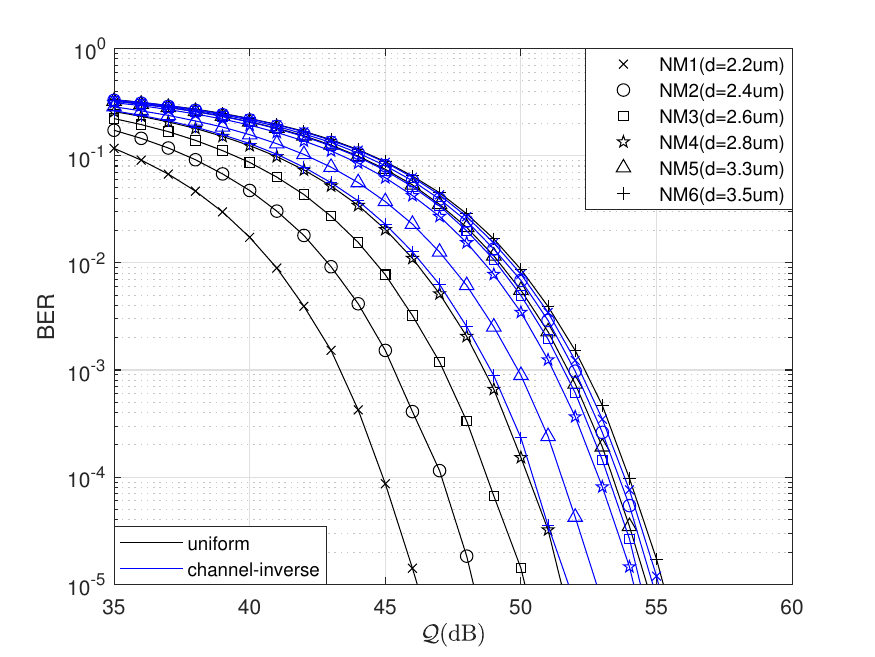,width=0.45\linewidth}}
  \caption{Comparison of BER performance of the MoCDMA systems with the parameters of $T=6\times10^{-2}s$, $L=10$, Walsh codes of $N=32$, supporting $6$ NMs as specified in Table.~\ref{tabel:parameter}.}
  \label{fig:BER_SQb}
\end{figure*}

In the context of Fig.~\ref{fig:BER_SQb}, the settings are the same as that for Fig.~\ref{fig:BER_KQb}, expect that in Fig.~\ref{fig:BER_SQb}, the  Walsh codes of length $N=32$ are employed. Compared to the results in Fig.~\ref{fig:BER_KQb} utilizing MLSs, employing Walsh codes can improve the BER performance of MoCDMA systems for both the emission methods and also for both the ZF- and MMSE-detectors, thanks to the orthogonality of Walsh codes. However, it is worth noting that the assignment of Walsh codes to NMs has a considerable impact on the BER performance of MoCDMA systems. This is because the locations of `-1's and `1's in the assigned Walsh codes result in different ICI, which hence affect the BER performance. Specifically, the Walsh codes with `1's occupying the first half and `-1's occupying the other half of the sequence are in favor of boosting the performance. By contrast, the Walsh codes having `-1' and `1' evenly distributed result in relatively large ICI. Thus, two Walsh code assignment methods are implemented in Fig.~\ref{fig:BER_SQb}, namely the best-to-closest (BTC) and best-to-furthest (BTF). Specifically, according to the locations of `-1's and `1's in a sequence, BTC assigns the best sequence to the NM closest to AP, the second best to the second closest, and so on; while BTF carries out the assignment in the way opposite to BTC, i.e., the best sequence to the NM furthest away from AP, and so on. By comparing Fig.~\ref{fig:BER_SQb}(a)(b) with (c)(d), it can be shown that the BTF scheme enhances the reliability of the NMs further away from AP at the cost of slightly degraded performance of the NMs closer to AP, when the uniform emission is employed. When the channel-inverse emission is employed, the BTF scheme makes the error performance of different NMs be distributed not as dense as that of the cases considered before in Fig.~\ref{fig:BER_SQb}(a)(b). Furthermore, the NM furthest away from AP achieves the best BER performance, while the one closest to AP obtains the worst BER performance. The reason for the above observation is that the best Walsh code not only boosts the signal power of the desired NM, but also generates severe interference on the other NMs, especially when this kind of Walsh codes are assigned to the furthest NMs, whose concentration tails can keep relatively large values for long time. On the other hand, the worst Walsh code with evenly distributed `-1's and `1's results in severe ICI to the desired NM, if it is assigned to the NM closest to AP, whose concentration decreases quickly. In the case of channel-inverses emission, the above influences of BTF on the BER performance can also be visualized.        

\begin{figure*}[htb!]
  \centering
    \subfigure[ZF-detector (BTC)]{\epsfig{figure=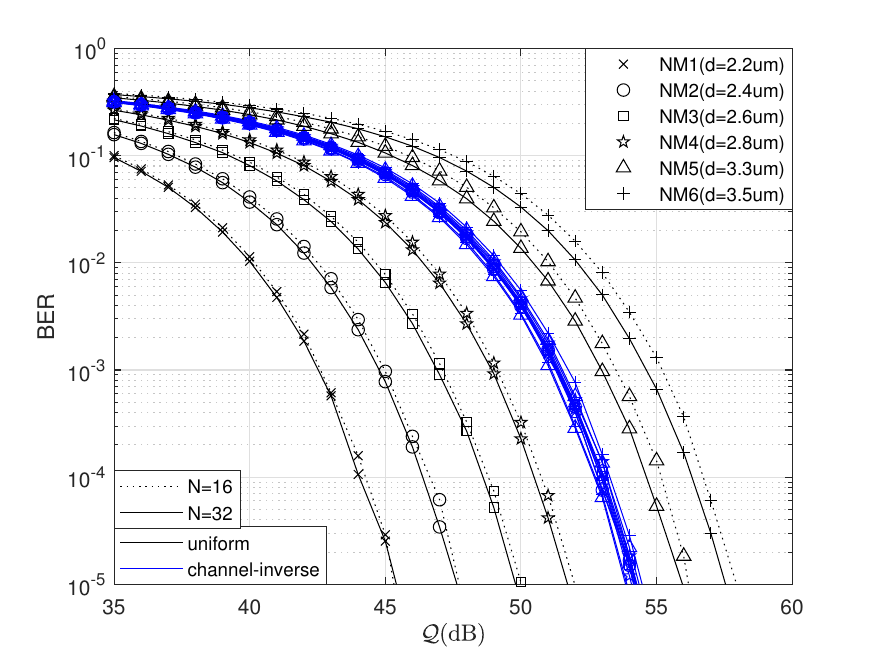,width=0.49\linewidth}} 
    \subfigure[MMSE-detector (BTC)]{\epsfig{figure=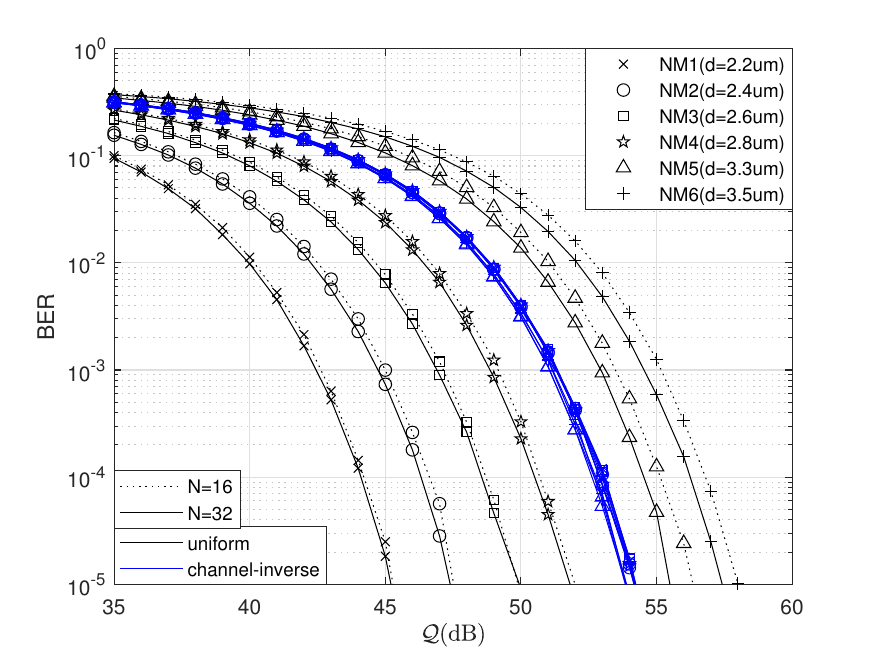,width=0.49\linewidth}}
  \caption{Impact of different spreading factors on the BER performance of the MoCDMA systems employing the Walsh codes assigned by following the BTC schemes, when the parameters are $T=6\times10^{-2}s$, $L=10$, and supporting $6$ NMs specified in Table.~\ref{tabel:parameter}.}
 \label{fig:BER_NQb}
\end{figure*}

As there are only 6 preferred MLSs of length $N=31$, which can support only upto $K=6$ NMs in a MoCDMA DMC system. By contrast, the Walsh codes of length $N=32$ can support more NMs. Therefore, in Fig.~\ref{fig:BER_NQb}, we investigate the impact of the spreading factor $N$ of the Walsh codes assigned based on the BTC scheme on the BER performance of the MoCDMA systems employing respectively the ZF- and MMSE-detectors cooperating. Compared to the results shown in Fig.~\ref{fig:BER_SQb}(a) and (b), Fig.~\ref{fig:BER_NQb} demonstrates that the decrease of spreading factor from $N=32$ to $16$ does not yield explicit degradation on the BER performance of the MoCDMA systems. It is well-known that in the conventional RdCDMA systems, the BER performance is usually explicitly dependent on the value of $K/N$. However, as shown in Fig.~\ref{fig:BER_NQb}, for both the detectors and also for both the emission methods, the BER performance is only slightly dependent on the spreading factor $N$. The reason behind is that in MoCDMA, the noise power is coupled with the number of NMs simultaneously supported and the interference generated by all NMs, regardless of what spreading factor is used. The noise power increases with the increase of the number of NMs, and furthermore, the noise power is unable to be reduced solely by the spreading and de-spreading operations. Therefore, we can be inferred that in MoCDMA systems, the BER performance is more dependent on the number of NMs, instead of the spreading factor. 

\section{Conclusion}\label{Conclusion}

A MoCDMA system relying on two types of molecules for supporting multiple NMs to transmit binary information to one AP was designed and studied. In our MoCDMA systems employing BMoSK, although the background noise is still signal-dependent, it is stationary and its statistics is not related to the specific information transmitted, which is beneficial to receiver design.  Owing to this, it was shown that the low-complexity MF, ZF, and MMSE detection schemes are similarly effective as that in the conventional RdCDMA systems. Our MoCDMA scheme assumes that NMs have different transmission distances from AP. Correspondingly, the uniform and channel-inverse based molecule-emission schemes were proposed and compared. The performance results illustrate that the channel-inverse emission is capable of improving the reliability of the NMs relatively `far away' from AP. This enables the detection of the NMs having different distances from AP to achieve a similar BER performance and, consequently, mitigate the near-far problem. Furthermore, our studies show that the MF-, ZF-, and MMSE detectors are all low-complexity detection schemes. The ZF- and MMSE-detectors are capable of achieving more reliable detection than the MF-detector, although both of them require slightly higher computations than the MF-detector. Hence, these detectors provide promising options for MoCDMA systems to attain a good trade-off between computational complexity and information reliability.



\end{document}